# Inelastic effects in electron tunneling through water layers

Michael Galperin and Abraham Nitzan
School of Chemistry, Tel Aviv University, Tel Aviv, 69978, Israel

February 28, 2001

## Abstract

Calculations of tunneling matrix elements associated with electron transfer through molecular environments are usually done for given frozen nuclear configurations; the underlying assumption being that nuclear motions are slow relative to the timescale of a tunneling event. This paper examines this issue for the case of electron tunneling through water. The motivation for this study is a recent calculation [Peskin et al, *J. Chem. Phys.* 111, 7558 (1999)] that indicates that electron tunneling through water may be enhanced by tunneling resonances in the range of ~1eV below the vacuum barrier, and finds that the lifetimes of such resonances are in the 10fs range, same order as OH stretch periods. Our calculation is based on the absorbing-boundary-conditions-Green's-function (ABCGF) method and proceeds in two steps. First we consider the effect of a single symmetric OH-stretch mode on electron tunneling in an otherwise frozen water environment and establish that the inelastic tunneling probability is small enough to justify an approach based on perturbation theory limited to single phonon transitions. Next we note that on the short timescale of a tunneling event, even under resonance conditions, water nuclear dynamics may be represented in the instantaneous normal modes picture. We generalize the ABCGF method to take into account low order inelastic scattering from a continuum of such harmonic normal modes. We find that near resonance the total inelastic transmission probability is of the same order as the elastic one, and may lead to an additional ~20-40% enhancement of the overall transmission in the range of up to 1eV below the vacuum barrier. The absolute energy exchange is small, of the order of 1% of the incident electron energy. Surprisingly, we find that the main contribution to the inelastic transmission is associated with energy transfer into the rotational-librational range of the water instantaneous normal mode spectrum.

## 1. Introduction

Electron tunneling through water is an important element in all electron transfer processes involving hydrated solutes, and in many processes that occur in water based electrochemistry. In the standard theory of electron transfer[1-3] the main role played by the solvent is to affect energy matching between donor and acceptor levels by thermal fluctuations. Continuum dielectric theory provides a reasonable description of this effect, however it obviously cannot describe the effect of solvent structure on the electronic coupling itself. In a recent series of articles [4-11] (see also [12]) we have addressed the problem using a molecular level description of the water environment: Water layer structures were sampled from equilibrium MD simulations and the electron transmission properties of these structures were analyzed numerically using a suitable pseudopotential to describe the electron-water interaction and applying the absorbing boundary condition Green function method to compute electron transmission probabilities. This numerical approach has yielded a comprehensive picture of the interplay between the electron transmission probabilities and the corresponding water structure. In particular, these studies have indicated that one possible source for lowering the effective barrier to tunneling in water is the possible involvement of transient water structures that support resonance tunneling. It was found that such resonances exist within ~ 1 eV below the bare vacuum barrier, and it was possible to correlate these resonances with particular water nanostructures. Furthermore, the lifetimes of these resonances was estimated to be in the order of 10fs. The involvement of resonance tunneling in electron transmission through water has been suggested also by other workers.[13] [14,15]

In the spirit of most calculations of electronic coupling elements in electron transfer theory and of earlier dielectric continuum models that disregard the water structure altogether, we have used in our previous studies of electron transmission probabilities static water structures. This *static medium approximation* is based on the assumption of short duration of the tunneling process, i.e., it is assumed that any transmission event is completed before substantial nuclear motion takes place. The computation of the transmission probability can therefore be done for individual static water configurations sampled from an equilibrium ensemble, and the results averaged over this ensemble. In turns this implies that electron transmission through water is essentially an elastic process. Indeed, experimental studies of photoemission through thin water films adsorbed on metals indicate that inelastic processes associated with the water nuclear motion contribute relatively weakly at energies of order ~1eV.[16,17] Numerical simulations of sub-excitation electron transmission through 1-4 water monolayers adsorbed on Pt(1,1,1)[18] are in agreement with this observation.[22] Theoretical calculations of inelastic tunneling[23] similarly show that sufficiently far from resonance the overall transmission is only weakly affected by inelastic processes.

The discovery[11] of transient resonance supporting structures, essentially molecular size cavities, in room temperature water, and the finding that the typical lifetimes of the corresponding resonances are of the same order as vibrational periods of intramolecular water nuclear motions cast some doubt on this static medium approximation. The lifetimes of these cavities are generally long relative to the lifetimes of the corresponding resonances,[24] so from this point of view a static picture is valid. However, the extent to which inelastic processes contribute to the transmission probability has to be reassessed. This is the purpose of the work presented in this paper.

The model used in the present calculation is similar to that used by us in earlier studies, and is briefly described in Section 2. In section 3 we relax the static medium approximation in an artificial way, keeping all nuclear coordinates in the water layer static except for one OH stretch mode associated with a single water molecule. We compute the inelastic transmission probabilities for processes that leave this mode in different vibrational states and conclude that inelastic tunneling is dominated by single phonon transitions. In Section 4 we calculate the 1-phonon inelastic electron transmission probability through water, invoking the short duration of the tunneling event by describing water motion in the instantaneous normal mode picture. As expected we find that inelastic tunneling is important close to resonance energies, however the overall energy exchange is small. As discussed in the concluding Section 5,imilarly the overall transmission probability is still approximated well by the static medium approximation.

## 2. The Model System

We use the same Pt(100)-$H_2O$-Pt(100) model system and the same electron-water pseudo-potential that were used in our previous calculations:[4-11] a polarizable flexible simple point charge (PFSPC) potential for the water-water interaction; Pt(100) surface with water-Pt interaction taken from the work of Spohr and Heinzinger;[25,26] and an electron-water pseudo-potential based on that developed by Barnett et al,[27] modified[7] to include the many body interaction associated with the water electronic polarizability of water. The electron-metal pseudo-potential is represented by a square barrier of height 5eV. The total potential experienced by the electron is assumed to be a superposition of these electron-water and electron-metal pseudopotentials. See Ref. 10 for more details. The studies described below are done for a distance 10Å between the Pt walls, which accommodates 3 water monolayers at normal density. Water configurations are prepared by classical MD evolution at 300K, using minimum image periodic boundary conditions in the directions parallel to the metal walls. Overall, 197 water molecules are contained in the simulation cell of dimensions 23.5×23.5×10Å, implying water density 1gr/$cm^3$. The system was equilibrated for 200 ps before selected water configurations were studied. Following the equilibration period we have collected water configurations at intervals of 10ps.

For each (static) water configuration obtained in this way the (elastic) electron transmission probability is computed as follows: a potential field for the electron motion is obtained as a superposition of the square barrier representing the electron-metal interaction and the electron water pseudopotential. This potential is projected into a cubic grid of dimension 16×16×400 with corresponding grid spacing 2.7713×2.7713×0.4au in the ($x,y,z$) directions ($z$ is the tunneling direction). The electronic problem is solved on this 23.46×23.46×84.67Å grid using periodic boundary conditions in the $x$ and $y$ directions and absorbing boundary conditions in the tunneling ($z$) direction; see Ref. 11 for details.[28] The computed transmission probabilities show distinct but configuration dependent resonance tunneling features below the bare (5 eV) work function that were traced[11] to the existence of transient nano-cavities in the water structure. Typical lifetimes of these resonances were found to be in the 10fs range.

## 3. Transmission in the presence of a single OH-stretch mode

In order to gain insight on the behavior of inelastic electron tunneling in water we focus first on the dynamics of a single normal mode of water in the otherwise frozen water layer. Expecting that inelastic coupling will be most effective for the high frequency OH stretch modes, [29] we 'unfreeze' the dynamics of a single OH stretch mode of a single water molecule in the layer and consider the resulting inelastic contributions to the tunneling process. For isolated water molecule whose geometry in the HOH plane is defined in terms of a 2-dimensional cartesian coordinate system centered on the O atom, with the y axis coinciding with the bisector of the HOH angle $\alpha$, standard procedure[30] yields the relations (see Appendix A) between the OH symmetric stretch mode, q, and the deviations $\delta x_1$, $\delta y_1$ (for $H_{[1]}$); $\delta x_2$, $\delta y_2$, (for $H_{[2]}$); $\delta x_0$, $\delta y_0$, (for O) of the H and O atoms from their equilibrium positions. If we unfreeze this normal mode for a single water molecule and let it move against the background of all other molecules it will move in a force field that is affected by this background. In the calculation described below, whose purpose is only to gauge the magnitude of the inelastic effect, we disregard this fact. We therefore consider an Hamiltonian

$$H = H_{el} + H_q + V_{el-q} \qquad (1)$$

where $H_{el}$ is the Hamiltonian for the motion of the electron interacting with the frozen water configuration,

$$H_{el} = K + U_B + V_{el-W} \qquad (2)$$

and where $K$ is the electron kinetic energy operator, $U_B(\mathbf{r})$ is the bare barrier (taken as a rectangular barrier of height 5eV and width 10Å) and $V_{el-W}(\mathbf{r};\mathbf{R}^N)$ is the electron-frozen water pseudo-potential. Here $\mathbf{r}$ and $\mathbf{R}^N$ denote the electron position and the frozen water configuration, respectively, and $N$ is the number of the molecules. Also in Eq. (1) $H_q = (\dot{q}^2 + \omega^2 q^2)/2$ is the Hamiltonian for the normal mode under consideration and $V_{el-q}(\mathbf{r},q)$ is the electron-normal mode interaction. An explicit form for $V_{el-q}(\mathbf{r},q)$ can be derived from the electron-water pseudopotential by writing the latter in the form

$$V_{el-W}(\mathbf{r},\mathbf{R}^N) = V(\mathbf{r},\mathbf{R}^{N-1}, x_o + \delta x_o, y_o + \delta y_o, x_1 + \delta x_1, y_1 + \delta y_1, x_2 + \delta x_2, y_2 + \delta y_2) \qquad (3)$$

where $(\mathbf{R}^{N-1}, x_0, y_0, x_1, y_1, x_2, y_2)$ denotes the underlying frozen configuration of the N-water molecule system, and by expressing $\delta x_j, \delta y_j$ $(j = 0,1,2)$ in terms of $q$ using Eqs. (43)-(46) of Appendix A. The resulting interaction obviously depends on the position of the particular water molecule with the unfrozen stretch mode, and on its orientation relative to the tunneling direction, but the general conclusions reached below are independent of these detail.

The numerical evaluation of the transmission probability associated with the Hamiltonian (1) is executed as a multichannel scattering problem. The electron motion is described on a spatial grid with absorbing potential set at the edges of the system in the tunneling direction as in Ref. 11, and the normal mode $q$ is described in the basis $\{|j\rangle\}$ of eigenfunctions of the free harmonic oscillator, defining the different channels. Because of its high frequency this mode starts out in the ground state, and inelastic tunneling therefore involves excitations of this mode to higher levels. Only open channels, i.e. with final nuclear energy not exceeding the total incoming electron energy,[31] should be included in the calculation. There is a finite number $n$ of such channels, with $n$ of order 5-10. Denoting the total number of grid points by $N_G$, and the number of vibrational states by $n$, the Hamiltonian $H$ in this representation is a matrix of order $n \times N_G$ that can be viewed as an $n \times n$ block matrix where each block is of order $N_G$. The diagonal blocks correspond to elastic motion of the electron in the different channels defined by the quantum states of the oscillator, while the non-diagonal blocks define for each grid point the coupling $V_{vv'}(\mathbf{r})=\langle v|V_{el-q}(\mathbf{r},q)|v'\rangle$ between the different channels $v,v'=0,...,n-1$. In the same representation the wavefunction of the electron-oscillator system is a column consisting of $n$ subcolumns of order $N_G$. Each subcolumn corresponds to the electron wavefunction in the corresponding channel.

We focus on the one-to-all transmission problem, where the electron is incident normal to the water layer and the total tunneling probability to exit in all final directions is calculated. The initial state that corresponds to the incoming electron is $\Phi_{in,v=0} = \phi_{in}(\mathbf{r})|v=0\rangle \sim e^{ikz}|v=0\rangle$ ($z$ is the transmission direction), with energy $E_{in} = \hbar^2 k^2 / 2m_e$. In the grid/oscillator basis the column vector that represents this state contains zeros everywhere except in the first subcolumn ($v=0$), which contains the grid representation of the wavefunction exp(*ikz*). The relevant absorbing-boundary-conditions Green's operator is

$$G = \frac{1}{E - H + i\varepsilon} \tag{4}$$

where $\varepsilon(z) = \varepsilon_{in}(z) + \varepsilon_{out}(z)$ is the sum of absorbing potentials defined at the incoming and outgoing edges of the numerical grid. The total transmission probability is obtained from[32,33] (see also Appendix B)

$$P = \frac{2}{\hbar} \langle \phi_{in} | \varepsilon_{in} G(E_{in}) \; \varepsilon_{out} G(E_{in}) \varepsilon_{in} | \phi_{in} \rangle \tag{5}$$

and the transmission probability into individual channels is given by

$$P_v = \frac{2}{\hbar} \langle \psi_v | \varepsilon_{out} | \psi_v \rangle \tag{6}$$

where $|\psi_v\rangle$ is the *v*-th block of scattering state supervector $|\psi\rangle = iG(E_{in})\varepsilon_{in} |\phi_{in}\rangle$.

Figure 1 displays results obtained from this calculation for one particular water configuration sampled from an equilibrium ensemble at 300K. Two cases with different water molecules from the middle water layer with "thawed" symmetric OH stretch were considered. In one case ("*a*") the thawed molecule is chosen such that one of its OH bonds is approximately parallel to the tunneling direction. In the other ("*b*") a molecule with the molecular plane perpendicular to the tunneling direction is chosen. The top full line in Fig. 1 shows the total transmission probability as a function of incident electron energy for both cases. This line coincides with the elastic transmission components in both cases because the inelastic contributions are extremely small. Results for inelastic transmission probabilities with transitions into levels *j*=1,2 of the oscillator are given by the lower full lines for case *a* and by the lower dashed lines for case *b*. The important observation is that, per mode, the inelastic contribution to the transmission probability in water are small, with the transmission probability for the 0→1 transition smaller by about five orders of magnitudes than its elastic counterpart. Figure 2 compares similar results for the 0→1 transition for water, $D_2O$ and $T_2O$. We find, as expected, that the inelastic coupling is stronger for the lighter isotope.

Obviously, the example shown here is highly artificial and the results will vary with water configuration and with the chosen water molecule. Still, we found the qualitative behavior and the order of magnitude of the results obtained in this way are characteristic of the model studied. We may conclude that under the conditions considered, inelastic tunneling is weak, implying that the effect of water nuclear motion can be treated as a small perturbation. This will be the starting point of the many-mode calculation described in the following Section.

## 4. Transmission through water described by instantaneous normal modes

The results of the previous section indicate that the interaction of the tunneling electron with a single water mode can be taken as a weak perturbation. The excitation of a single quantum of the symmetric OH stretch mode was seen to be a very low probability event and the probability to induce more than a single phonon transition during the tunneling event is extremely small even near the transmission resonance. Still, since the electron interacts with a large number of modes (the normalization volume used in the calculation of Sect. 3 corresponds to an electron interacting with a sample of ~200 water molecules, i.e. with ~2000 nuclear degrees of freedom), the total inelastic contribution may be appreciable. In this section we provide an approximate estimate of this contribution.

The calculation described below relies on the fact, discussed above, that the timescale for the interaction of the tunneling electron with the water environment is short. The upper limit for this timescale is provided by the lifetimes of the transmission resonances, estimated to be in the order of 10fs.[11] In addition a calculation of the traversal time for electron tunneling through water layers similar to those used in the present calculation [34] yield times of the order ≤1fs away from the resonance energy, and again times of order ~10fs near resonance. These short times are consistent with the low probability for inelastic transitions observed in the calculation of Section 3 Still, as discussed in the introduction, nuclear motion on this timescale cannot be disregarded altogether. Instead we use previous work that indicate that on timescales shorter than ~100fs the nuclear dynamics of different solvents, including water, can be reliably described in the instantaneous normal mode picture.[35-38] In this approach each static liquid configuration sampled from the equilibrium ensemble is used as a reference for expanding the nuclear potential up to second order in the deviations of the nuclear coordinates from this reference configuration. Diagonalizing the Hessian matrix associated with the second order derivatives of the nuclear potential about this reference point then yields a set of 'instantaneous normal modes' (INMs) whose dynamics describes the short time evolution of the liquid about that configuration. In a similar spirit we assume that the inelastic contribution to electron transmission through a thin (a few molecular monolayers) water layer can be accounted for by the interaction of the tunneling electron with the instantaneous normal modes associated with the configuration encountered at each tunneling event. Previous studies have indicated that taking into account only stable (real frequency) INMs does not lead to appreciable errors in the very short time dynamics of the fluid considered, and provides a better representation of the exact fluid dynamics than an all-mode treatment at longer times.[36-38]

Here we follow the same practice and represent the water dynamics using only the subset of stable INMs. In the discussion below we refer to these modes as 'phonons', keeping in mind the limited applicability of this term in the present context.

The inelastic contribution to the transmission probability is computed as the average of results obtained in this way over configurations sampled from an equilibrium ensemble of layer configurations. Furthermore, based on the results of Section 3 we assume that the inelastic process is dominated by single phonon transitions that may be evaluated using perturbation theory. This use of perturbation theory to evaluate the inelastic contribution to electron transmission through water, combined with the description of the water nuclear dynamics as the motion of non-interacting normal modes, is what makes this problem tractable.

For any given water configuration $\mathbf{R}_0$ our starting point is the Hamiltonian

$$H = H_{el}(\mathbf{R}_0) + H_{ph}(\mathbf{R}_0) + V_{el-ph}(\mathbf{r}, \mathbf{R}_0) \tag{7}$$

$$H_{el} = K_{el} + U_B(\mathbf{r}) + V_{el-W}(\mathbf{r}, \mathbf{R}_0) \tag{8}$$

where $\mathbf{r}$ is the electron coordinate, $H_{el}$ is the Hamiltonian associated with the motion of the electron across the static metal-water-metal junction, $H_{ph}$ is the phonon Hamiltonian (see Appendix B, Eq. (51)) and $V_{el-ph}$ is the electron-phonon interaction that we discuss below. $H_{el}$ may be further written as a sum of electron kinetic energy $K$, the bare barrier $U_B$ that is modeled here as a rectangular barrier of height 5eV, and the electron-water pseudopotential $V_{el-W}$ discussed in Section 2.

For a given instantaneous water configuration, $V_{el-ph}$ is obtained from expanding the electron-water interaction in powers of the deviations of the water-atoms positions from this instantaneous configuration. Denoting the water configuration close to a reference configuration $\mathbf{R}_0$ by $\mathbf{R} = \mathbf{R}_0 + \sum_\alpha C_\alpha q_\alpha$ where $\{q_\alpha\}$ are the instantaneous normal modes coordinates associated with that reference configuration (see Appendix B), the electron-water interaction is expanded up to linear order in $\{q_\alpha\}$

$$V_{el-W}(\mathbf{r},\mathbf{R}) = V_{el-W}(\mathbf{r},\mathbf{R}_0) + \sum_\alpha U_\alpha(\mathbf{r},\mathbf{R}_0) q_\alpha \qquad (9)$$

with $U_\alpha(\mathbf{r},\mathbf{R}_0) = (\partial V_{el-W}(\mathbf{r},\mathbf{R}_0)/\partial \mathbf{R}_0) \cdot \partial \mathbf{R}_0/\partial q_\alpha$, so that in Eq. (7)

$$V_{el-ph} = \sum_\alpha U_\alpha(\mathbf{r},\mathbf{R}_0) \cdot q_\alpha \qquad (10)$$

In Eqs. (9) and (10) we have assumed that the order linear in the INM coordinates is sufficient. This is consistent with our expectation that the inelastic contribution to the tunneling process is dominated by single phonon transitions. One more approximation was made in the actual application of this procedure: As already discussed in Section 2 the electron-water pseudopotential used in the present work is based on that developed by Barnett et al,[27] modified to incorporate the many-body nature of the effect of the water electronic polarizability. We have found that including this aspect of the electron-water interaction is essential for getting the correct effective barrier for electron tunneling through water. However, the effect of these terms on the dynamics of inelastic tunneling is expected to be small relative to the stronger shorter-range parts of the pseudopotential, and are disregarded in the evaluation the $U_\alpha$ terms. This simplifies the calculation of these terms considerably.

Next we consider the needed scattering formalism. The Hamiltonian (7) describes an electron that is moving as a free particle to the right and left of a target, i.e. $U$, $V_{el-W}$ and $V_{el-ph}$ are taken to vanish for $\mathbf{r}$ outside the target. This target is a water layer superimposed on the vacuum barrier. In the process under discussion the electron tunnels through this barrier, while interacting with the INMs bath localized in the barrier region.

We denote by $|l\rangle = |\phi_l\rangle|\chi_l\rangle$ and $|r\rangle = |\phi_r\rangle|\chi_r\rangle$ the initial and final states of the electron-water system, where $|\phi_l\rangle$ and $|\phi_r\rangle$ are respectively incoming and outgoing states on the left and right sides of the barrier, while $|\chi_l\rangle$ and $|\chi_r\rangle$ are the initial and final nuclear states of the water target. The corresponding energies are $E_{l,r} = E_{l,r}^{el} + E_{l,r}^{ph}$ and the relevant $S$ matrix element is

$$S_{rl} = -2\pi i T_{rl}(E_l)\delta(E_r - E_l) \tag{11}$$

For the scattering geometry considered here the corresponding $T$ matrix element may be written in the form

$$T_{rl}(E) = \langle r|H_{RM}G(E)H_{LM}|l\rangle \tag{12}$$

by dividing space into three regions: The right ($R$) and left ($L$) regions of free particle motion, and a target region ($M$) that encloses the molecular layer. With this in mind we will sometimes write the electronic Hamiltonian in the form

$$H_{el} = H_{el}^0 + H_{LM} + H_{RM} \tag{13}$$

where $H_{el}^0$ represents the electronic motion in the disconnected subspaces R, L and M and where $H_{RM}$ and $H_{LM}$ are the couplings between these subspaces. Eq. (12) assumes that no direct coupling exists between the $R$ and $L$ regions. In the spatial grid representation used $H_{RM}$ and $H_{LM}$ originate from the kinetic energy operator, and direct $R$-$L$ coupling may indeed be disregarded if the target encompasses more than a few grid spacings in the tunneling direction. From the structure of Eq. (12) it follows that the Green function $G$ enters only through its projection on the M subspace, $G_{MM}(E) = [E - H_{MM} - \Sigma(E)]^{-1}$, where $H_{MM}$ is the projection of the Hamiltonian (7) on the M subspace and where $\Sigma$ is the corresponding self-energy. $G_{MM}$ is a matrix of the order of the number of grid points in the M region. Note that, since $V_{el-ph}(\mathbf{r},\mathbf{R}_0)$ vanishes for $\mathbf{r}$ outside the molecular layer, $\Sigma$ is an operator in the electronic space only.

To first order in the electron-phonon interaction we may write

$$G(E) = G^0(E) + G^0(E)V^{el-ph}G^0(E) \tag{14}$$

$$G^0(E) = \left[E - H_{el} - H_{ph} + i\eta\right]^{-1} \quad ; \quad (\eta \to 0) \tag{15}$$

From the discussion above it is clear that we need only the projected part of $G^0$, $G^0_{MM}(E) = \left[E - (H_{el})_{MM} - H_{ph} - \Sigma(E)\right]^{-1}$. Obviously, $G^0_{MM}$ can couple only between states $|l\rangle$ and $|r\rangle$ with $|\chi_l\rangle = |\chi_r\rangle$ and, since $H_{ph}$ commutes with the electronic operators $H_{el}$ and $\Sigma$, it can be replaced in (15) by the corresponding phonon energy. For this reason we will also encounter below the electronic Green's function

$$G_{el}(E) = \left[E - H_{el} + i\eta\right]^{-1} \tag{16}$$

and its M-projected part

$$(G_{el})_{MM}(E) = \left[E - (H_{el})_{MM} - \Sigma(E)\right]^{-1} \tag{17}$$

With these preliminaries we are now ready to calculate the matrix element $T_{rl}(E)$ of Eq. (12). Below we use interchangeably the notations $|l\rangle = |l_{el} l_{ph}\rangle = |\phi_l \chi_l\rangle$ (and similarly for $r$). Using Eq. (14) we may write this coupling as a sum of elastic and inelastic components

$$T_{lr} = T_{lr}^{elas} + T_{lr}^{inelas} \tag{18}$$

$$\begin{aligned} T_{lr}^{elas}(E) &= \langle \phi_l \chi_l | H_{LM} G^0(E) H_{RM} | \phi_r \chi_r \rangle \\ &= \langle \phi_l | H_{LM} G_{el}(E - E^{ph}) H_{RM} | \phi_r \rangle \delta_{l_{ph} r_{ph}} \end{aligned} \tag{19}$$

$$\begin{aligned} T_{lr}^{inelas}(E) &= \langle \phi_l \chi_l | H_{LM} G^0(E) V^{el-ph} G^0(E) H_{RM} | \phi_r \chi_r \rangle = \\ &= \sum_\alpha \langle \phi_l | H_{LM} G_{el}(E_l^{el}) U_\alpha(\mathbf{r}) G_{el}(E_r^{el}) H_{RM} | \phi_r \rangle \langle \chi_l | q_\alpha | \chi_r \rangle \end{aligned} \tag{20}$$

In Eq. (19) Kronecker $\delta$ term denotes that the initial ($l_{ph}$) and final ($r_{ph}$) states of the phonons are identical in the elastic process, and $E^{ph} = E_l^{ph} = E_r^{ph}$ is the corresponding phonon energy. In Eq. (20) $E_j^{el} = E - E_j^{ph}$ ; $j = l, r$, and in each term of the $\alpha$ summation $E_l^{el}$ and $E_r^{el}$ differ from each other by one quantum of the mode $\alpha$. In addition, the structures of Eqs. (19) and (20) imply that only the M-projected parts of the Green's operators in these equations, i.e., $G_{MM}^{(0)}$ and $(G_{el})_{MM}$ respectively, should be considered.

For simplicity of notation we drop here and below this projected notation, keeping in mind that the Green's functions appearing below are all projected onto the M subspace.

We will focus on a particular 'one-to-all' transmission probability whereupon the electron is incident from the left in the direction normal to the water layer and the sum over all final states is considered. The one-to-all transmission probability is given by

$$P_l = \frac{2\pi}{\hbar} \sum_r |T_{lr}|^2 \, \delta(E_l - E_r) \tag{21}$$

In the presence of phonons this should be averaged over their initial distribution and summed over their final states to yield the total averaged transmission probability.

$$\langle P_l \rangle = \frac{2\pi}{\hbar} \sum_{r_{el}} \frac{1}{Z_L} \sum_{l_{ph}} e^{-\beta E_l^{ph}} \sum_{r_{ph}} |T_{lr}|^2 \, \delta(E_l^{el} + E_l^{ph} - E_r^{el} - E_r^{ph}) \tag{22}$$

Here $Z_L$ is the phonon partition function in the initial state. Note that the summations in (22) are done for a given incoming state $l_{el}$ of the electron, and therefore at constant $E_l^{el}$. Also, in what follows we shall omit the explicit average $\langle \, \rangle$, writing merely $P_l$ for the averaged transmission probability when the meaning is clear from the text.

Consider first the elastic tunneling component. The sum over final phonon states eliminates the $\delta_{l_{ph}, r_{ph}}$ term in the square of Eq. (19) and the average over initial phonon states is trivial: $Z_L^{-1} \sum_{l_{ph}} \exp(-\beta E_l^{ph}) = 1,$ so effectively the $|T|^2$ term becomes purely electronic. It leads (see Appendix C) to the following expresion for the one-to-all transmission probability

$$P_l^{elastic}(E_l^{el}) = \frac{1}{\hbar} \langle \phi_l | H_{ML} G_{el} \Gamma^R G_{el} H_{ML} | \phi_l \rangle \tag{23}$$

where $\Gamma^{(R)} = i \left[ \Sigma^{(R)} - \Sigma^{(R)} \right]$. In Eqs. (21)-(23) the wavefunction is normalized to unit flux in the incoming wave

$$\phi_E(x, y, z) \xrightarrow{z \to -\infty} \sqrt{\frac{m}{k_z \hbar}} e^{i\mathbf{k}_z \cdot \mathbf{z}} f(x, y), \quad E = \hbar^2 k_z^2 / 2m + E_{xy} \tag{24}$$

where $E_{xy}$ is the energy in the direction normal to $z$ and $f(x,y)$ is normalized to 1. In the absorbing boundary conditions (ABC) Green's function method the boundaries separating the left and right free electron regions $L$ and $R$ from the 'molecular' region $M$ are taken far enough from the target to allow the replacement of the self energy martix Σ by simple position dependent imaginary potential terms $i\varepsilon_L(\mathbf{r})$ and $i\varepsilon_R(\mathbf{r})$ that rise smoothly towards the boundaries of the M system and insure the absorption of outgoing waves at these boundaries. This leads to (see Appendix C)

$$P_l^{elastic}(E_l^{el}) = \frac{2}{\hbar}\langle\phi_l|\varepsilon_L G_{el}(E_l^{el})\varepsilon_R G_{el}(E_l^{el})\varepsilon_L|\phi_l\rangle \tag{25}$$

where $G_{el}(E) = \left[E - H_{el} + i(\varepsilon_L + \varepsilon_R)\right]^{-1}$. The numerical evaluation of this expression requires (a) evaluating the Hamiltonian matrix on the grid, and (b) evaluating the operation of the corresponding Green's operator on the vector $|\phi_l\rangle$. In our implementation 7$^{th}$ order finite-differencing representation is used to evaluate the kinetic energy operator on the grid. This results in a sparse matrix representation of the Hamiltonian, suggesting the applicability of Krylov space based iterative methods[39] for such calculations. In the present work we have used the PETSc package.[40]

In Appendix C we develop an equivalent procedure for evaluating inelastic transmission probabilities. The result equivalent to (25) is

$$P_l^{inelast}(E_l^{el}) = \sum_{\alpha=1}^{N}\left[g_\alpha^+(E_l^{el})h_\alpha^+ + g_\alpha^-(E_l^{el})h_\alpha^-\right] \tag{26}$$

where

$$g_\alpha^\pm(E_l^{el}) \equiv \frac{2}{\hbar}\langle\phi_l|\varepsilon_L G_{el}(E_l^{el})U_\alpha(\mathbf{r})G_{el}(E_l^{el}\pm\hbar\omega_\alpha)\varepsilon_R G_{el}(E_l^{el}\pm\hbar\omega_\alpha)U_\alpha(\mathbf{r})G_{el}(E_l^{el})\varepsilon_L|\phi_l\rangle \tag{27}$$

$$h_\alpha^+ = h^+(\omega_\alpha) \equiv \frac{\hbar}{2m_\alpha\omega_\alpha}\bar{n}(\omega_\alpha); \quad h_\alpha^- = h^-(\omega_\alpha) \equiv \frac{\hbar}{2m_\alpha\omega_\alpha}(1+\bar{n}(\omega_\alpha)) \tag{28}$$

with $\bar{n}(\omega) = (\exp(\beta\hbar\omega) - 1)^{-1}$. Note that in the position representation $U_\alpha(\mathbf{r})$ is real and diagonal, i.e. $U_\alpha(\mathbf{r}) = U_\alpha(\mathbf{r})$. Also, since we used mass weighted coordinates, $m_\alpha = 1$.

In order to compute the matrix element in Eq. (27) one needs to evaluate the vector $G_{el}(E_l^{el}\pm\hbar\omega_\alpha)U_\alpha(\mathbf{r})G_{el}(E_l^{el})\varepsilon_L|\phi_l\rangle$. This involves two operations of Green's matrices on a vector that are carried out as outlined below Eq. (25). Obviously, an exact calculation of the sum (26) involves too many operations of this kind to be practical. Instead we resort to a coarse-graining approximation. First rewrite Eq. (26) in the form

$$P_l^{inelast}(E_l^{el}) = \int_0^\infty d\omega \rho(\omega)\left[g^+(E_l^{el},\omega)h^+(\omega) + g^-(E_l^{el},\omega)h^-(\omega)\right] \qquad (29)$$

where

$$g^\pm(E_l^{el},\omega) = \rho^{-1}(\omega)\sum_\alpha g_\alpha^\pm(E_l^{el})\delta(\omega-\omega_\alpha) \qquad (30)$$

and where $\rho(\omega) = \sum_\alpha \delta(\omega-\omega_\alpha)$ is the density of phonon modes. The approximation is to use the following expression for $g^\pm(E_l^{el},\omega)$

$$g^\pm(E_l^{el},\omega) = \frac{2}{\hbar}\langle\phi_l^{el}|\varepsilon_L G_{el}(E_l^{el})U(\mathbf{r},\omega)G_{el}(E_l^{el}\pm\hbar\omega)\varepsilon_R G_{el}(E_l^{el}\pm\hbar\omega)U(\mathbf{r},\omega)G_{el}(E_l^{el})\varepsilon_L|\phi_l^{el}\rangle \qquad (31)$$

where $U^2(\mathbf{r},\omega) = \rho^{-1}(\omega)\sum_\alpha U_\alpha^2 \delta(\omega-\omega_\alpha)$. Numerically, an approximation to $U(\mathbf{r},\omega)$ is obtained by dividing the $\omega$ axis into a finite number of segments $\{i\}$ of sizes $\Delta\omega_i$ and taking the average of $U_\alpha^2$ over all modes in a segment, i.e.

$$U(\mathbf{r},\omega) = \left(\frac{1}{N(\omega)}\sum_{\omega-\Delta\omega/2\leq\omega_\alpha<\omega+\Delta\omega/2} U_\alpha^2(\mathbf{r})\right)^{1/2} \qquad (32)$$

where $N(\omega)=\rho(\omega)\Delta\omega$ is number of modes in the segment considered.

Eq. (29) yields, under our approximations, the total probability for inelastic tunneling. Obviously, the term involving $g^+$ in this equation corresponds to processes in which the electron gained energy, while the term involving $g^-$ is associated with processes in which the electron lost energy during the transmission. These terms are integrals of the corresponding differential transmission probabilities

$$\mathcal{P}_l^{inelast}(E_l^{el},\pm\hbar\omega) = \rho(\omega)g^\pm(E_l^{el},\omega)h^\pm(\omega) \qquad (33)$$

To summarize, our calculation proceeds along the following steps.

(a) Generate an equilibrium trajectory of the water layer confined between the two Pt(100) surfaces and use it to sample a desired number of water layer configurations.

(b) For any such water configuration and for an appropriate spatial grid we evaluate the grid representation of the inverse Green function $E - H_{el} + i(\varepsilon_L + \varepsilon_R)$.

(b) For a given water configuration find the corresponding set of instantaneous normal modes using the potential and procedure described in Appendix B. We proceed with the subset of stable modes of real frequency.

(c) The electron-water pseudo-potential is expanded in the normal modes about the given water configuration. The linear terms in this expansion yield the coupling parameters $U_\alpha(\mathbf{r})$ according to Eqs. (9)-(10).

(d) The coarse-grained representation $U(\mathbf{r},\omega)$ of $U_\alpha(\mathbf{r})$ is obtained using Eq. (32).

(e) For the given incoming state $|\phi_l^{el}\rangle$, represented as a grid vector, we calculate the vector $G^{el}(E_l^{el} \pm \hbar\omega) U(\mathbf{r},\omega) G^{el}(E_l^{el}) \varepsilon_L |\phi_l^{el}\rangle$ and use it to obtain $g^{\pm}(E_l^{el},\omega)$, Eq. (31). The latter is then used in Eqs. (29) and (33).

## 4. Results and discussion

Figure 3a shows the density of instantaneous normal modes for the system described in Section 2: three monolayers of normal water and of $D_2O$ confined between two Pt surfaces at 300K. Figure 3b show similar results for "bulk" water[41] at 300K and at 60K, together with those obtained for the water layer. These results are averaged over 20 water configurations sampled from the corresponding equilibrium trajectories. The results for bulk water are in close agreement with those obtained previously by other workers.[35,36] The results obtained for the water layer are very similar to those obtained for the bulk system of the same temperature except at the low frequency regime $\omega \leq 500 \text{cm}^{-1}$. In this regime the normal mode spectrum of the confined layer is seen to shift to somewhat higher frequencies and at the same time the number of imaginary frequency modes is reduced relative to the corresponding spectral features of bulk water. Both observations reflect the increased local binding in the confined layer. The absence of appreciable effects in the higher frequency regime may be understood by noting the frequencies associated with the O-Pt stretch and the O-Pt-Pt bend of a single water molecule adsorbed on Pt(100) (based on the potential of Ref. [25,26] are $97 \text{cm}^{-1}$ and $504 \text{cm}^{-1}$ respectively.

The elastic and integrated inelastic one-to-all (incident perpendicular direction) transmission probabilities for one randomly chosen configuration of the 3-monolayer water film are shown in Fig. 4. Figure 4a shows the elastic and inelastic components of the transmission probabilities as well as the total transmission probability - their sum - as functions of the incident electron energy. Fig. 4b shows the elastic transmission probability as well as the corresponding inelastic components obtained for identical $H_2O$, $D_2O$ and $T_2O$ layers. The resonance enhancement of the transmission probability near 4.4eV (for the studied configuration) is seen to be accompanied by a substantial increase in the inelastic component. It is important to keep in mind the transient nature and the variability of the water layer structures that support such resonances. The ratio $P_{inelast}/P_{elast}$ between the elastic and (integrated) inelastic components of the transmission is shown for several different water configurations in Fig. 5.

The results of Fig. 4b indicate that the importance of inelastic electron transmission through isotopically substituted water layers decreases in the order H>D>T, as expected from the relative masses.[42] It is also interesting to examine the relative importance of different phonon-frequency ranges in affecting inelastic transmission. Time-scale arguments would suggest that the electron couples more efficiently to the higher frequency modes associated with the intramolecular stretching and bending vibrations, however additional factors should be taken into account. First, because the terms $h^{-}(\omega)$ (for electron energy loss) and $h^{+}(\omega)$ (for electron energy gain) in Eq. (29) are proportional to $\bar{n}(\omega)+1$ and to $\bar{n}(\omega)$, respectively, tunneling accompanied by electron energy loss is more important than tunneling with energy gain, particularly for high frequency modes with $\hbar\omega > k_B T$. On the other hand, the transmission probability for an energy loss process is reduced relative to the corresponding elastic process because of the larger barrier encountered effectively in the latter process. For the combination of these reasons the low frequency regime associated with water rotations and librations is more important in affecting inelastic tunneling than the higher frequency intramolecular regime. This is seen in Figure 6 where the net differential energy loss spectrum,

$$\frac{d(\Delta E)}{d\omega} = \frac{d}{d\omega}\left((\Delta E)_{loss} - (\Delta E)_{gain}\right) = \hbar\omega\left(\mathcal{P}_l^{inelast}(-\hbar\omega) - \mathcal{P}_l^{inelast}(+\hbar\omega)\right) \quad (34)$$

is displayed as a function of the phonon frequency ω. Shown are the spectral distributions of the net energy loss computed for identical configurations of $H_2O$ and $D_2O$ layers for two values of the incident energy $E_l^{el}$: at resonance (4.4eV for the configuration used), Fig. 6a, and substantially below resonance (3.5eV), Fig. 6b. Figures 7a and 7b show for the same configuration and the same incident energies the two components, $(d/d\omega)(\Delta E)_{loss}$ (on the negative frequency axis) and $(d/d\omega)(\Delta E)_{gain}$ (on the positive frequency side) as functions of the phonon frequency.

Finally, Fig. 8 depicts the total energy loss, as well as some of its spectral components, for the same water configuration used in Figs. 6 and 7. The prominence of the low frequency regime ($\omega$<1000cm$^{-1}$) in affecting inelastic tunneling is clearly seen. It should be emphasized that, while this spectral regime contains also contributions from oxygen translational modes, the dominance of hydrogenic motions is evident from the strong sensitivity to hydrogen isotope substitution seen in Figs. 6 and 7. The actual net energy loss is quite small, substantially less than 1% of the incident electron energy.

## 5. Conclusions

In this work we have assessed the relative importance of inelastic effects in electron tunneling through water layers. It was prompted by previous studies that indicated that electron tunneling through water is enhanced by resonaces associated with transient water structures characterized by molecular cavities in the water structure. A generalization of the absorbing boundary conditions Green's function methodology makes it possible to compute first order (single phonon) corrections to the elastic tunneling probability. We have combined this with an instantaneous normal mode representation of the short time-scale nuclear dynamics of the water layer in order to compute the effect of inelastic tunneling on the water transmission properties.

The results described above indicate that indeed near resonance energies inelastic tunneling cannot be disregarded. For particular configurations we find inelastic currents that exceed the elastic component. A rough average over several configurations using data such as in Fig. 5 leads to a modest increase of the total tunneling probability by 20-40% in the range of ~1eV below the vacuum barrier due to inelastic contributions. In the deep tunneling regime (incident energy lower than 1eV below the vacuum barrier) inelastic

19contributions can be disregarded as far as their effect on the overall transmission is concerned.

We have also examined the separate contributions of different parts of the instantaneous water phonon spectrum to the inelastic tunneling process. We have found that relatively low frequency modes involving hydrogen rotations and librations are the main nuclear motions affecting inelastic electron tunneling through water.

Finally, we should keep in mind that the calculation described above has focused only on transmission properties of the water layer. In a calculation of the actual tunneling current the transmission probability should be summed over all initial and final energies with proper account given to the Fermi population factors that determine the availability of electrons with the corresponding initial energy and the accessibility of final energy states. The differential transmission probability, Eq. (33), which is function of both initial and final electron energies, provides in principle the input for such a calculation.

## Appendix A

We denote by 0, 1 and 2 the oxygen atom and the two hydrogen atoms, respectively. Using standard procedure, the relationship between $q$, the symmetric OH-stretch mode, and the deviations $\delta x_1$, $\delta y_1$ (for $H_{[1]}$); $\delta x_2$, $\delta y_2$, (for $H_{[2]}$); $\delta x_0$, $\delta y_0$, (for O) of the H and O atoms from their equilibrium positions in the molecular plane $xy$ can be written as follows. Define

$$\kappa_{11} = \frac{1}{m_H}\left( k_{OH} \sin^2\alpha + \frac{2k_\alpha}{r^2}\cos^2\alpha \right) \tag{35}$$

$$\kappa_{22} = \frac{1}{m_H}\left(1 + \frac{2m_H}{m_O}\right)\left( k_{OH} \cos^2\alpha + \frac{2k_\alpha}{r^2}\sin^2\alpha \right) \tag{36}$$

$$\kappa_{12} = \kappa_{21} = \frac{1}{2m_H}\sqrt{1 + \frac{2m_H}{m_O}}\left( k_{OH} - \frac{2k_\alpha}{r^2} \right)\sin 2\alpha \tag{37}$$

$$m_1 = 2m_H \tag{38}$$

$$m_2 = 2m_H\left(1 + \frac{2m_H}{m_O}\right) \tag{39}$$

$$\omega = \left[1/2\left(\kappa_{11} + \kappa_{22} + \sqrt{(\kappa_{11} + \kappa_{22})^2 - 4(\kappa_{11}\kappa_{22} - \kappa_{12}^2)}\right)\right]^{1/2} \tag{40}$$

$$c = \frac{\omega^2 - \kappa_{22}}{\kappa_{12}} \tag{41}$$

$$N = \sqrt{1 + c^2} \tag{42}$$

where $m_H$ and $m_O$ are the atomic masses of the H and O atoms, respectively, $r$ is the equilibrium OH bond length, $\alpha$ is the equilibrium HOH angle, $k_{OH}$ is the force constant for the OH bond and $k_\alpha$ is the force constant for the angle $\alpha$ (so that the potential for harmonic deviations of the water configuration from equilibrium is $(1/2)k_{OH}(\delta r_{01})^2 + (1/2)k_{OH}(\delta r_{02})^2 + (1/2)k_\alpha(\delta\alpha)^2$). Then

$$\delta x_0 = 0 \tag{43}$$

$$\delta y_0 = -\frac{2m_H}{m_O\sqrt{m_2}}\frac{1}{N}q \tag{44}$$

$$\delta x_1 = -\delta x_2 = \frac{C}{N\sqrt{m_1}}q \tag{45}$$

$$\delta y_1 = \delta y_2 = \frac{1}{N\sqrt{m_2}}q \tag{46}$$

In the preliminary study discussed in Sect. 3 we take the Hamiltonian for the thawed normal mode q to be the same as in a free water molecule, $H_q = (\dot{q}^2 + \omega^2 q^2)/2$.

**Appendix B. Water potential and instantaneous normal modes (INMs).**

A set of instantaneous normal modes associated with a given water configuration is obtained by expanding the water potential to second order in the deviations from the given configuration

$$V_W(\mathbf{R}) - V_W(\mathbf{R}_0) = -\mathbf{F}(\mathbf{R}_0)\cdot(\mathbf{R} - \mathbf{R}_0) + \frac{1}{2}(\mathbf{R} - \mathbf{R}_0)\cdot\mathbf{D}(\mathbf{R}_0)\cdot(\mathbf{R} - \mathbf{R}_0) \tag{47}$$

where **r** are mass weighted coordinates and where

$$F_j(\mathbf{R}) = -\frac{\partial V_W(\mathbf{R})}{\partial R_j} \tag{48}$$

$$D_{jk}(\mathbf{R}) = \frac{\partial^2 V_W(\mathbf{R})}{\partial R_j \partial R_k} \tag{49}$$

and diagonalizing the Hessian matrix **D**. Denoting by **M** the corresponding transformation matrix, the squared normal modes frequencies, $\omega_\alpha^2$, are the elements of the diagonal matrix $\mathbf{\Omega}^2 = \mathbf{MDM}$ and the corresponding shifted coordinates are $\mathbf{q} = \mathbf{M} \cdot (\mathbf{R} - \mathbf{R}_0) - (\mathbf{\Omega}^2)^{-1}\mathbf{f}$, where $\mathbf{f} = \mathbf{M} \cdot \mathbf{F}$ is the transformed force. The short time evolution about the configuration is then determined by the Hamiltonian

$$H_{ph} = \sum_\alpha \left( (1/2)\dot{q}_\alpha^2 + (1/2)\omega_\alpha^2 q_\alpha^2 \right), \tag{50}$$

where $\omega_\alpha^2$ are the diagonal elements of $\mathbf{\Omega}^2$. In general, some of the INM's frequencies are imaginary and in the present tunneling calculation we have taken into account only stable modes with real frequencies that we refer to as 'phonons'. The starting point of the calculation outlined in Sect. 4 is the second quantization representation of the Hamiltonian (50)

$$H_{ph} = \sum_\alpha \hbar\omega_\alpha \left( a_\alpha a_\alpha + \frac{1}{2} \right) \tag{51}$$

where $a_\alpha$ and $a_\alpha$ are phonon creation and annihilation operators and where the sum is over stable INMs.

The nuclear potential for the water system is

$$V_W = V_{Wb} + V_{Pt-W} \tag{52}$$

Here $V_{Wb}$ is the bulk water potential while $V_{Pt-W}$ is the water-platinum wall interaction. For the former we use the flexible simple point charge (FSPC) model[9,43,44] where the potential is written as a sum of intermolecular and intramolecular contributions

$$V_{Wb} = V_W^{inter} + V_W^{intra} \tag{53}$$

The forms of these potentials used in the present calculation are as follows:

(a) The intermolecular water term is a sum of pair interactions between water atoms not belonging to the same molecule, as in the SPC water model

$$V_W^{inter} = \sum_{i<j} v_W^{inter}(i,j) \quad \text{(sums over water pairs)} \tag{54}$$

$$v_W^{\mathrm{inter}}(i,j) = \frac{q_H^2}{R_{H_{1i}H_{1j}}} + \frac{q_H^2}{R_{H_{2i}H_{2j}}} + \frac{q_H^2}{R_{H_{1i}H_{2j}}} + \frac{q_H^2}{R_{H_{2i}H_{1j}}} + \frac{q_O q_H}{R_{O_i H_{1j}}} + \frac{q_O q_H}{R_{O_i H_{2j}}} + \frac{q_O q_H}{R_{H_{1i}O_j}} + \frac{q_O q_H}{R_{H_{2i}O_j}}$$

$$+ \frac{q_O^2}{R_{O_i O_j}} + 4\varepsilon \left(\frac{\sigma}{R_{O_i O_j}}\right)^6 \left[\left(\frac{\sigma}{R_{O_i O_j}}\right)^6 - 1\right]$$

(55)

where, e.g. $R_{H_{1i}H_{1j}} = |\mathbf{R}_{H_{1i}} - \mathbf{R}_{H_{1j}}|$, with $\mathbf{R}_{H_{1i}}$ being the position of the hydrogen "1" on molecule $i$. The parameters used in Eq. (55) are $q_O$ = -0.82a.u., $q_H$ = 0.41a.u., $\sigma$ and $\varepsilon$=0.1554 kcal/mol.

(b) The intramolecular water term is a sum over all water molecules

$$V_W^{\mathrm{intra}} = \frac{1}{2}\sum_i v_{H_2O}^{\mathrm{intra}}(i) \tag{56}$$

of terms of the form[45]

$$v_W^{\mathrm{intra}}(i) = K_1\left(d_{1i}^2 + d_{2i}^2\right) + 2K_2 d_{1i}d_{2i} + K_3 d_{ai}^2 + 2K_4(d_{1i}+d_{2i})d_{ai} + K_5\left(d_{1i}^3 + d_{2i}^3\right)$$

$$+ K_6(d_{1i}+d_{2i})d_{1i}d_{2i} + K_7\left(d_{1i}^2+d_{2i}^2\right)d_{ai} + K_8 d_{1i}d_{2i}d_{ai} + K_9(d_{1i}+d_{2i})d_{ai}^2 + K_{10}d_{ai}^3 \quad (57)$$

$$+ K_{11}\left(d_{1i}^4 + d_{2i}^4\right) + K_{12} d_{1i}^3 d_{2i}^3 + K_{13} d_{1i}^2 d_{2i}^2 + K_{14}\left(d_{1i}^2+d_{2i}^2\right)d_{a_i}^2 + K_{15} d_{1i}d_{2i}d_{ai}^2$$

where

$$d_{1i} = \left|\mathbf{R}_i^{H_1} - \mathbf{R}_i^O\right| - R_{eq}$$
$$d_{2i} = \left|\mathbf{R}_i^{H_2} - \mathbf{R}_i^O\right| - R_{eq} \tag{58}$$
$$d_{a_i} = \mathbf{R}_{eq}\left\{\arccos\frac{\left(\mathbf{R}_i^{H_1} - \mathbf{R}_i^O\right)\cdot\left(\mathbf{R}_i^{H_2} - \mathbf{R}_i^O\right)}{\left|\mathbf{R}_i^{H_1} - \mathbf{R}_i^O\right|\cdot\left|\mathbf{R}_i^{H_2} - \mathbf{R}_i^O\right|} - \alpha_{eq}\right\}$$

The parameters used in Eqs. (57) and (58) are $K_i(i=1...3)$=8431.9269, -100.73624, 759.01307·$10^{-21}$J/Å$^2$, $K_i(i=4...10)$=262.1435, -9525.0947, -319.16137, 159.58416, -658.27683, 149.60711, -139.63423·$10^{-21}$J/Å$^3$ and $K_i(i=11...15)$=15359.793, 797.88605, 1296.5996, -1695.5426, -498.64404·$10^{-21}$J/Å$^4$, $R_{eq}$=1Å and $\alpha_{eq}$=109.5°.

(c) The water-Pt(100) interaction is taken from Refs. [25,26]. It takes the form

$$V_{Pt-H_2O} = \sum_{j\in Pt}\left\{\sum_{i\in O} v_{Pt-O}^{(i,j)} + \sum_{i\in H} v_{Pt-H}^{(i,j)}\right\} \tag{59}$$

with

$$v_{Pt-O}^{(i,j)} = \left[ae^{-\alpha r_{ij}} - be^{-\beta r_{ij}}\right]f(\rho_{ij}) + ce^{-\gamma r_{ij}}\left[1 - f(\rho_{ij})\right] \qquad (60)$$

$$R_{ij} = \left|\mathbf{R}_i - \mathbf{R}_j\right| \qquad (61)$$

$$\rho_{ij} = \left|(\mathbf{R}_i - \mathbf{R}_j) - \mathbf{n}\left((\mathbf{R}_i - \mathbf{R}_j)\cdot\mathbf{n}\right)\right| \qquad (62)$$

$$f(\rho) = \exp(-\delta\rho^2) \qquad (63)$$

where **n** is a unit vector normal to the Pt wall and pointing in the outwards (i.e. into the water) direction and $i$ and $j$ refer to the oxygen an Pt atoms respectively. Also

$$v_{Pt-H}^{(i,j)} = g\exp(-\eta r_{ij}) \qquad (64)$$

where $i$ and $j$ refer to hydrogen and Pt positions respectively. The parameters used in these equations are $a=1.8942\cdot10^{-16}$J, $b=1.8863\cdot10^{-16}$J, $c=10^{-13}$J, $\alpha=1.1004$Å$^{-1}$, $\beta=1.0966$Å$^{-1}$, $\gamma=5.3568$Å$^{-1}$, $\delta=0.5208$Å$^{-2}$, $g=1.7142\cdot10^{-19}$J and $\eta=1.2777$Å$^{-1}$. In the equations above energy is given in J while **R** and $\rho$ are expressed in Å.

## Appendix C

Here we provide some details pertaining to the derivations of Eqs. (23), (25) and (26)-(27). Consider first the elastic transmission rate out of electronic state $l$

$$k_l^{elast} = \frac{2\pi}{\hbar}\sum_r \left|\langle\phi_l|H_{LM}G_{el}(E_l^{el})H_{RM}|\phi_r\rangle\right|^2 \delta(E_l^{el} - E_r^{el}) \qquad (65)$$

Expanding in the basis of $m$ states that span the M subspace this can be written in the form

$$k_l^{elast} = \frac{2\pi}{\hbar}\sum_{m_1}\sum_{m_2}\sum_{m_3}\sum_{m_4}\sum_r \langle\phi_l|H_{LM}|m_1\rangle\left(G_{el}(E_l^{el})\right)_{m_1 m_2}\langle m_2|H_{RM}|\phi_r\rangle \\
\times \langle\phi_r|H_{RM}|m_3\rangle\left(G_{el}(E_l^{el})\right)_{m_3 m_4}\langle m_4|H_{LM}|\phi_l\rangle\delta(E_l^{el} - E_r^{el}) \qquad (66)$$

Next use the identity

$$\Gamma_{m_2,m_3}^R(E_l^{el}) \equiv i\left[\Sigma^R - \Sigma^R\right] = 2\pi\sum_r \langle m_2|H_{RM}|\phi_r\rangle\langle\phi_r|H_{RM}|m_3\rangle\delta(E_l^{el} - E_r^{el}) \qquad (67)$$

and do the formal sums over the $m$ states to get

$$k_l^{elast}(E_l^{el}) = \frac{1}{\hbar}\langle\phi_l|H_{LM}G_{el}(E_l^{el})\Gamma^R(E_l^{el})G_{el}(E_l^{el})H_{LM}|\phi_l\rangle \qquad (68)$$

This rate can be converted to transmission probability by dividing it by the incoming flux, $j_l = \hbar k_z / m$ where $m$ is the electron mass. Alternatively we may write the transmission probability as

$$P_l^{elast}(E_l^{el}) = \frac{1}{\hbar} \langle \phi_l | H_{LM} G_{el}(E_l^{el}) \Gamma^R(E_l^{el}) G_{el}(E_l^{el}) H_{LM} | \phi_l \rangle \tag{69}$$

provided that the wavefunction $\phi_l$ is normalized to unit incoming flux, i.e.

$$\phi_E(x,y,z) \xrightarrow{z \to -\infty} \sqrt{\frac{m}{k_z \hbar}} e^{i\mathbf{k}_z \cdot \mathbf{z}} f(x,y), \quad E = \hbar^2 k_z^2 / 2m + E_{xy} \tag{70}$$

($E_{xy}$ is the energy in the direction normal to $z$ and $f(x,y)$ is normalized to 1). In the absorbing boundary conditions (ABC) Green's function method the boundaries separating the left and right free electron regions $L$ and $R$ from the 'molecular' region $M$ are taken far enough from the target to allow the replacement of the self energy matrix $\Sigma$ by simple position dependent imaginary potential terms $i\varepsilon_L(\mathbf{r})$ and $i\varepsilon_R(\mathbf{r})$ that rise smoothly towards the boundaries of the M system and insure the absorption of outgoing waves at these boundaries. In this case $\Gamma^{(R)}(E_l^{el})$ in Eq. (69) may be replaced by $2\varepsilon_R$. A further simplification may be achieved if we use the fact that the incoming electronic state $\phi_l$ is localized in the L subspace and may be taken to vanish near the RM boundary. This implies that $\varepsilon_R |\phi_l\rangle = 0$ and $H_{RM}|\phi_l\rangle = 0$. Using these equalities we can write

$$\varepsilon_R G_{el} H_{LM} |\phi_l\rangle = \varepsilon_R G_{el} \left( [G_{el}]^{-1} + H_{LM} + H_{RM} \right) |\phi_l\rangle. \tag{71}$$

Together with $\left[ G_{el}(E_l^{el}) \right]^{-1} = E_l^{el} - H_0^{el} - H_{LM} - H_{RM} + i\varepsilon$ and $\left( E_l^{el} - H_0^{el} \right)|\phi_l\rangle = 0$ this yields

$$\varepsilon_R G_{el} H_{LM} |\phi_l\rangle = \varepsilon_R G_{el} \varepsilon_L |\phi_l^{el}\rangle \tag{72}$$

so that

$$P_l^{elastic}(E_l^{el}) = \frac{2}{\hbar} \langle \phi_l | \varepsilon_L G_{el}(E_l^{el}) \varepsilon_R G_{el}(E_l^{el}) \varepsilon_L | \phi_l \rangle \tag{73}$$

Consider next the inelastic transmission probability. Inserting Eq. (20) into (22) yields the inelastic rate

$$k_l^{inelast}(E_l^{el}) = \frac{2\pi}{\hbar}\sum_{r_{el}}\sum_{r_{ph}}\frac{1}{Z_L}\sum_{l_{ph}}e^{-\beta E_l^{ph}}\sum_{\alpha}\sum_{\alpha'}\langle\phi_l|H_{LM}G_{el}(E_l^{el})U_{\alpha'}(\mathbf{r})G_{el}(E_r^{el})H_{RM}|\phi_r\rangle$$

$$\times\langle\phi_r|H_{RM}G_{el}(E_r^{el})U_{\alpha}(\mathbf{r})G_{el}(E_l^{el})H_{LM}|\phi_l\rangle\langle\chi_l|q_{\alpha'}|\chi_r\rangle\langle\chi_r|q_{\alpha}|\chi_l\rangle\delta\left(E_r^{el}+E_r^{ph}-E_l^{el}-E_l^{ph}\right)$$

(74)

where $E_j^{el} = E_j - E_j^{ph}$; $j = l,r$ is the electronic energy. Again, a thermal averaging over the initial states $l_{ph}$ of the phonon bath and sums over the final states ($r_{el}, r_{ph}$) of the electron and bath are taken in (74). The summation and averaging over the phonon degrees of freedom can be done by noting that (a) only terms with $\alpha=\alpha'$ contribute in (74), and (b) in the term associated with a given mode $\alpha$, the vibrational states $|\chi_l\rangle$ and $|\chi_r\rangle$ (themselves products of single mode states) differ from each other by a change of one quantum in the state of mode $\alpha$, so that $E_r^{el} = E_l^{el} \pm \hbar\omega_{\alpha}$. This leads to

$$k_l^{inelast}(E_l^{el}) = \frac{2\pi}{\hbar}\times\sum_{\alpha}\sum_r\frac{\hbar}{2m_a\omega_{\alpha}}\{$$

$$\langle\phi_l|H_{LM}G_{el}(E_l^{el})U_{\alpha}(\mathbf{r})G_{el}(E_r^{el})H_{RM}|\phi_r\rangle\langle\phi_r|H_{RM}G_{el}(E_r^{el})U_{\alpha}(\mathbf{r})G_{el}(E_l^{el})H_{LM}|\phi_l\rangle \quad (75)$$

$$\times\left[\bar{n}_{\alpha}\delta(E_l^{el}+\hbar\omega_{\alpha}-E_r^{el})+(\bar{n}_{\alpha}+1)\delta(E_l^{el}-\hbar\omega_{\alpha}-E_r^{el})\right]\}$$

Note that since we used mass-weighted coordinates, $m_{\alpha}=1$ for all $\alpha$. Repeating the same procedure that was used to get Eq. (68) now yields

$$k_l^{inelast}(E_l^{el}) = \frac{2\pi}{\hbar}\times\sum_{\alpha}\frac{\hbar}{2m_a\omega_{\alpha}}\{$$

$$\bar{n}_{\alpha}\langle\phi_l|H_{LM}G_{el}(E_l^{el})U_{\alpha}(\mathbf{r})G_{el}(E_l+\hbar\omega_{\alpha})\Gamma^R(E_l+\hbar\omega_{\alpha})G_{el}(E_l+\hbar\omega_{\alpha})U_{\alpha}(\mathbf{r})G_{el}(E_l^{el})H_{LM}|\phi_l\rangle +$$

$$(\bar{n}_{\alpha}+1)\langle\phi_l|H_{LM}G_{el}(E_l^{el})U_{\alpha}(\mathbf{r})G_{el}(E_l-\hbar\omega_{\alpha})\Gamma^R(E_l-\hbar\omega_{\alpha})G_{el}(E_l-\hbar\omega_{\alpha})U_{\alpha}(\mathbf{r})G_{el}(E_l^{el})H_{LM}|\phi_l\rangle\}$$

(76)

Finally, using the the ABC-Green's function methodology and the arguments that lead to Eq. (73) we now get

$$P_l^{elastic}(E_l^{el}) = \sum_{\alpha=1}^{N}\left[g_{\alpha}^{+}(E_l^{el})h_{\alpha}^{+} + g_{\alpha}^{-}(E_l^{el})h_{\alpha}^{-}\right] \qquad (77)$$

where

$$g_{\alpha}^{\pm}(E_l^{el}) \equiv \frac{2}{\hbar}\langle\phi_l|\varepsilon_L G_{el}(E_l^{el})U_{\alpha}(\mathbf{r})G_{el}(E_l^{el}\pm\hbar\omega_{\alpha})\varepsilon_R G_{el}(E_l^{el}\pm\hbar\omega_{\alpha})U_{\alpha}(\mathbf{r})G_{el}(E_l^{el})\varepsilon_L|\phi_l\rangle$$

(78)

$$h_\alpha^+ \equiv \frac{\hbar}{2m\omega_\alpha}\bar{n}_\alpha; \quad h_\alpha^- \equiv \frac{\hbar}{2m\omega_\alpha}(1+\bar{n}_\alpha) \tag{79}$$

and where in Eq. (78) the functions $|\phi_l\rangle$ are normalized to unit flux, as in Eq. (70). Note that in our mass-weighted coordinate system $m_\alpha=1$ for all $\alpha$.

**Acknowledgements.** This research was supported by the Israel Ministry of Science and by the U.S.-Israel Binational Science Foundation. We thank Dr. Sivan Toledo for introducing us to and helping with the PETSc package.

**Figure captions**

Fig. 1. The transmission probability plotted against the incident electron energy for an electron interacting in the barrier with a single OH-stretch mode of one water molecule. Top full line - the total transmission probability that coincides on the scale shown with the elastic component. The lower full lines show, for a molecule in orientation "a" (see text), the inelastic transmission probabilities associated with one (upper line) and two (lower line) quantum excitations of the symmetric OH stretch vibration. The lower dashed lines show similar results for a molecule in orientation "b".

Fig. 2. Full line - same as the full line in Fig. 1. Dashed, dotted and dashed-dotted lines show the inelastic transmission probabilities with one-quantum excitation of the OH-stretch mode for H, D and T substituted water, respectively. The molecule is in configuration "a" (see text).

Fig. 3. (a) The density $\rho$ of instantaneous normal modes in a layer comprised of three monolayers of water molecules confined between two static Pt(100) surfaces, averaged over 20 configurations sampled from an equilibrium (T=300K) trajectory and shown for otherwise identical $H_2O$ (dashed line) and $D_2O$ (dotted line) layers. The usual convention of displaying unstable modes on the negative frequency axis is applied here. (b) Same for bulk water systems at 60K (full line) and 300K (dotted line) shown together with the result for a water layer (dashed line, same as the dashed in (a). The densities of modes shown are normalized to 1.

Fig. 4. (a) The elastic (full line) and the inelastic (dashed line) transmission probabilities as well as their sum - the total transmission probability (dotted line), plotted against the incoming electron energy (b) The elastic transmission probability (full line) and the inelastic transmission probabilities (dashed line for $H_2O$, dotted line for $D_2O$ and dashed-dotted line for $T_2O$) displayed against the incoming (perpendicular to the water layer) electron energy.

Fig. 5. The ratio $P_{inelast} / P_{elast}$ between the inelastic (integrated over all transmitted energies) and elastic components of the transmission probability calculated for different

instantaneous structures of a water layer consisting of 3 monolayers of water molecules confined between two Pt(100) surfaces.

Fig. 6. The net energy loss spectrum, Eq. (34), for an electron tunneling through a given water layer configuration (a) at resonance ($E_l^{el}$=4.4eV for the chosen configuration) and (b) substantially below resonance ($E_l^{el}$=3.5eV). Dashed line is for $H_2O$ and dotted line is for $D_2O$.

Fig. 7. The energy loss/gain spectrum for the same process depicted in Fig. 6 (a) at resonance ($E_l^{el}$=4.4eV) and (b) below resonance ($E_l^{el}$=3.5eV). The gain spectrum is shown on the positive $\omega$ axis (right vertical scale) while the loss spectrum is displayed on the negative $\omega$ side (left vertical scale). Dashed and dotted lines are for $H_2O$ and $D_2O$ respectively.

Fig. 8. The net energy loss computed for the same water ($H_2O$) configuration as used in Figs. 7 and 8, displayed as a function of the incident electron energy. Full line - the total net energy loss integrated over all phonon frequencies $\omega$. Dashed line - the energy loss integrated over the $\omega<1300\text{cm}^{-1}$ regime. dotted line - same for the range that spans the intramolecular bending frequency, $1500\text{cm}^{-1}<\omega<2000\text{cm}^{-1}$. Dashed dotted line - same for the intramolecular stretch frequencies range, $\omega>3000\text{cm}^{-1}$.

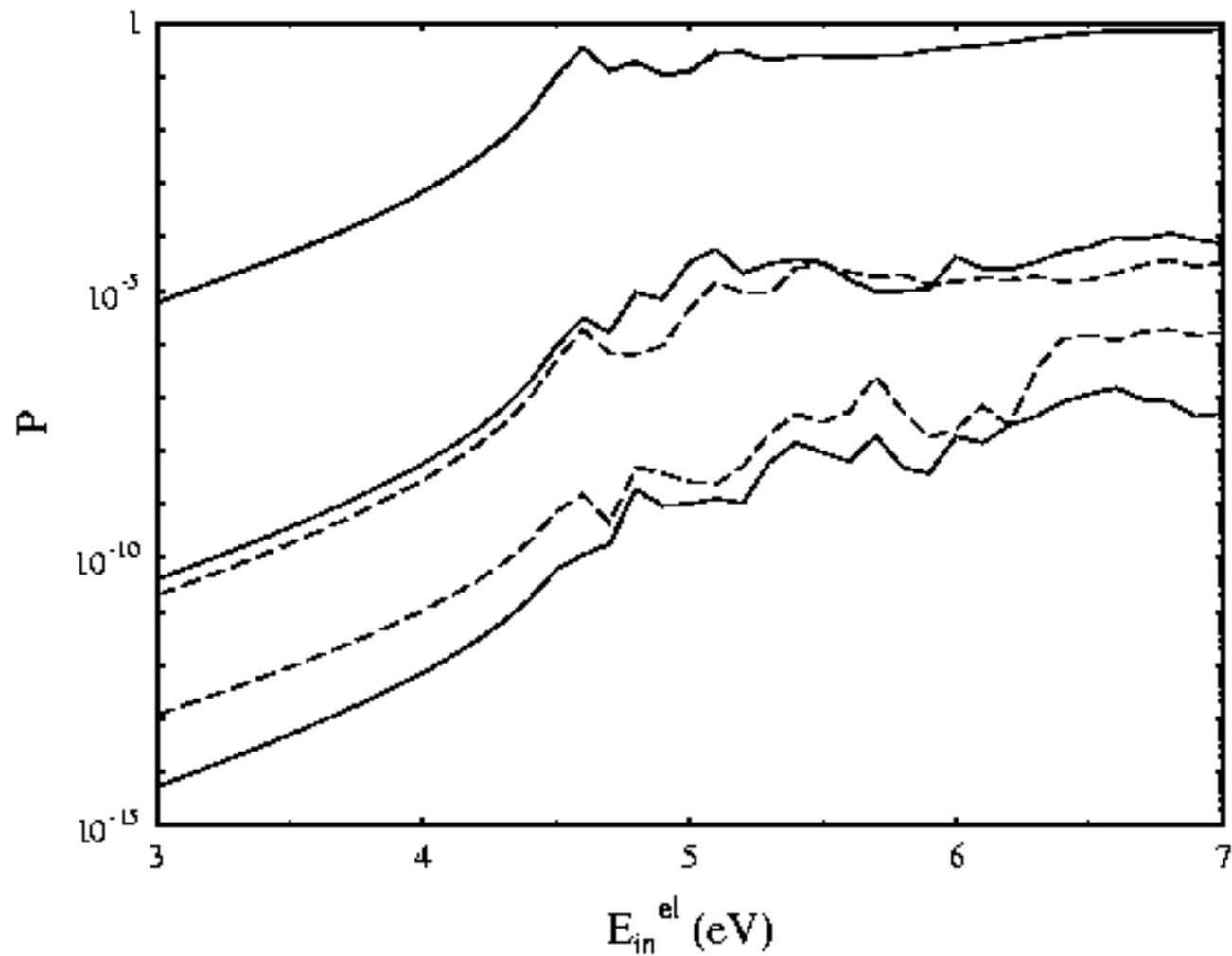

FIG. 1

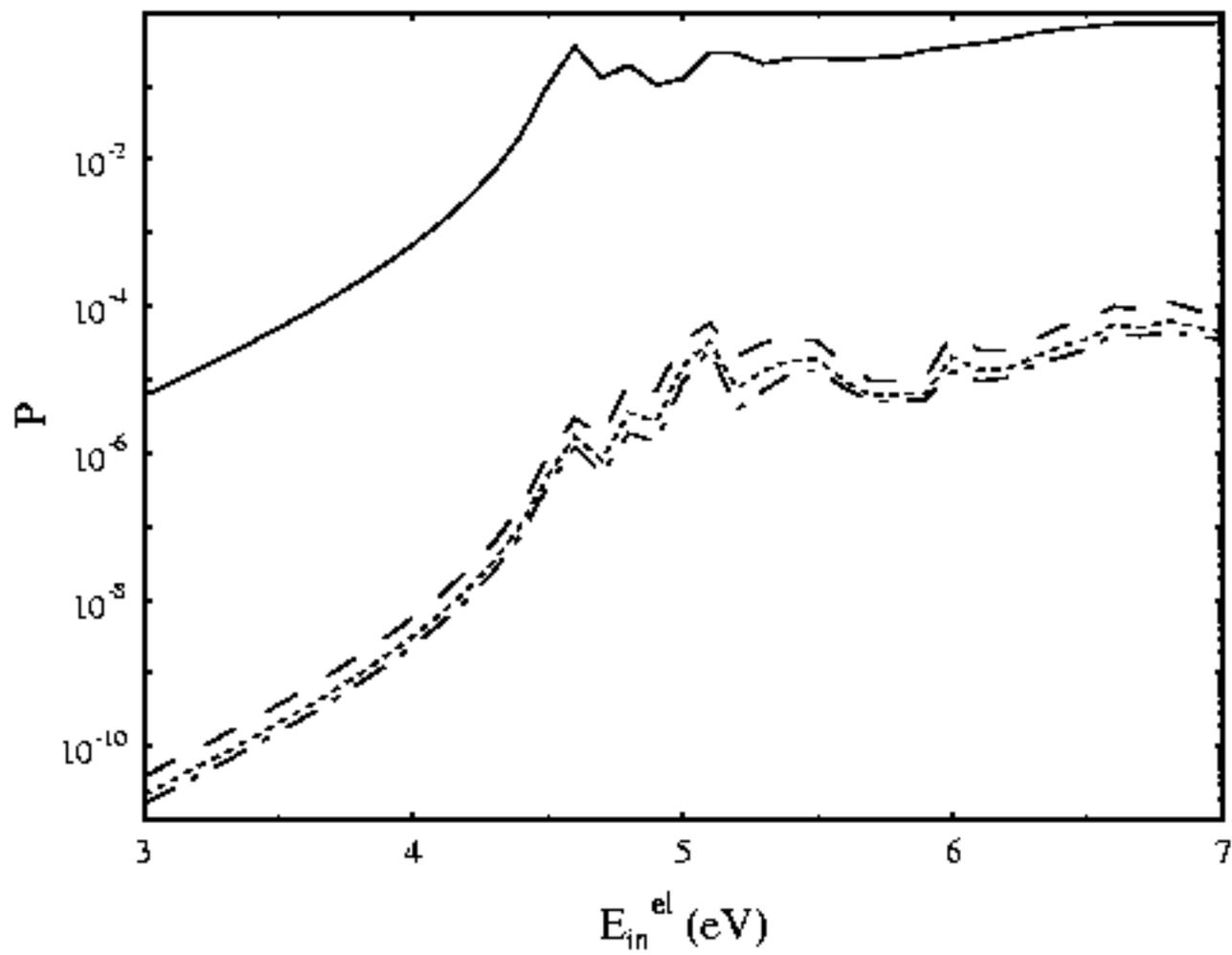

FIG. 2.

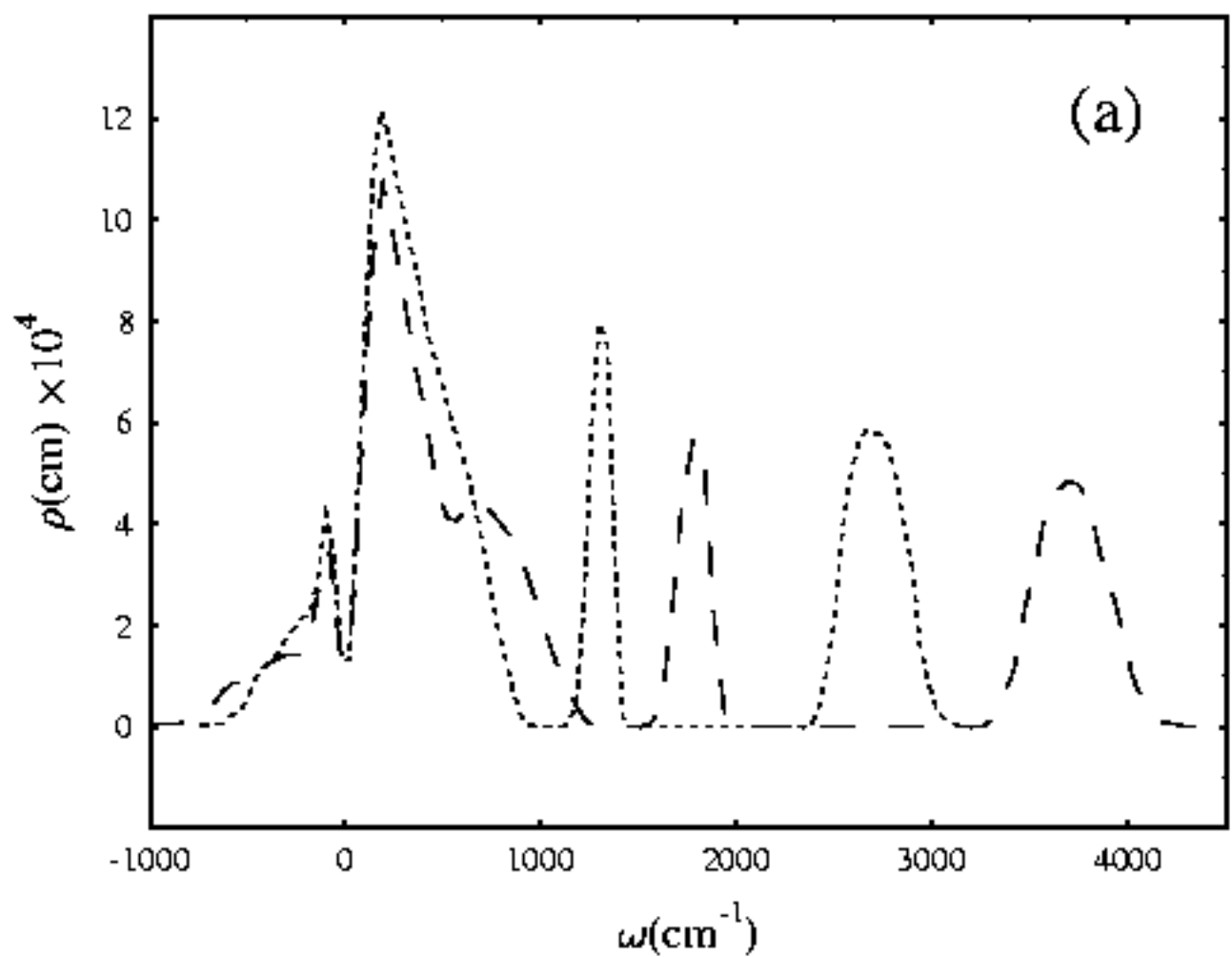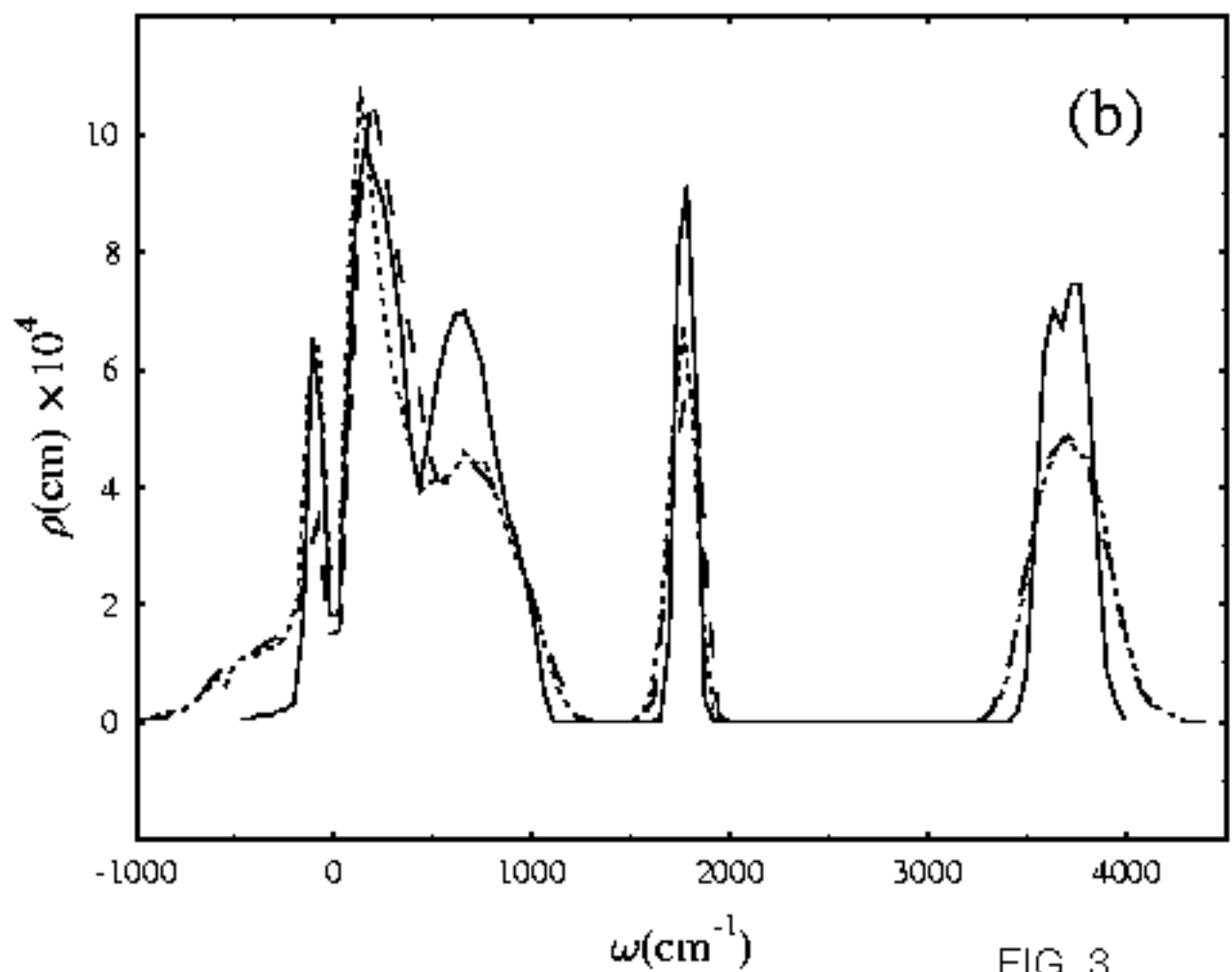

FIG. 3.

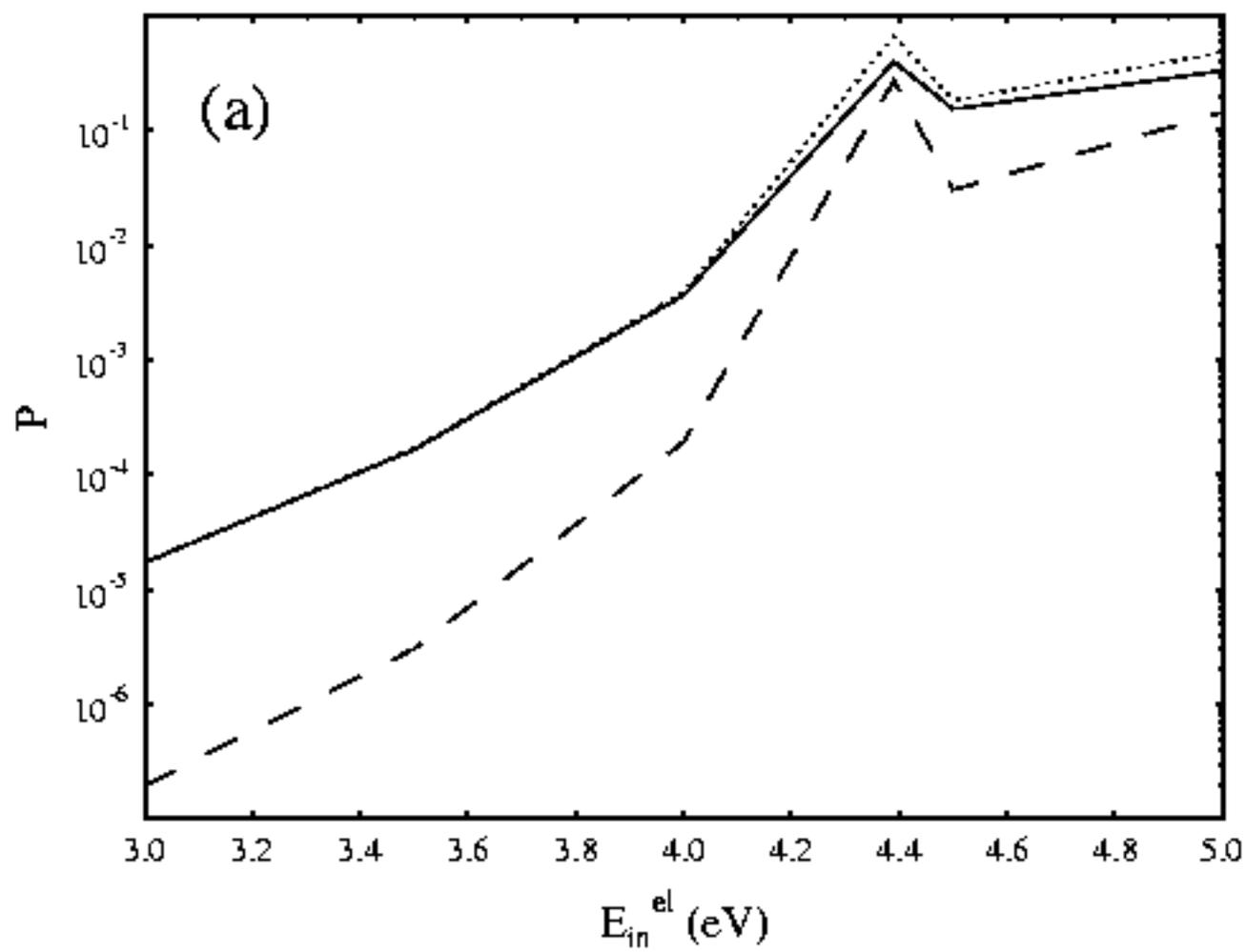

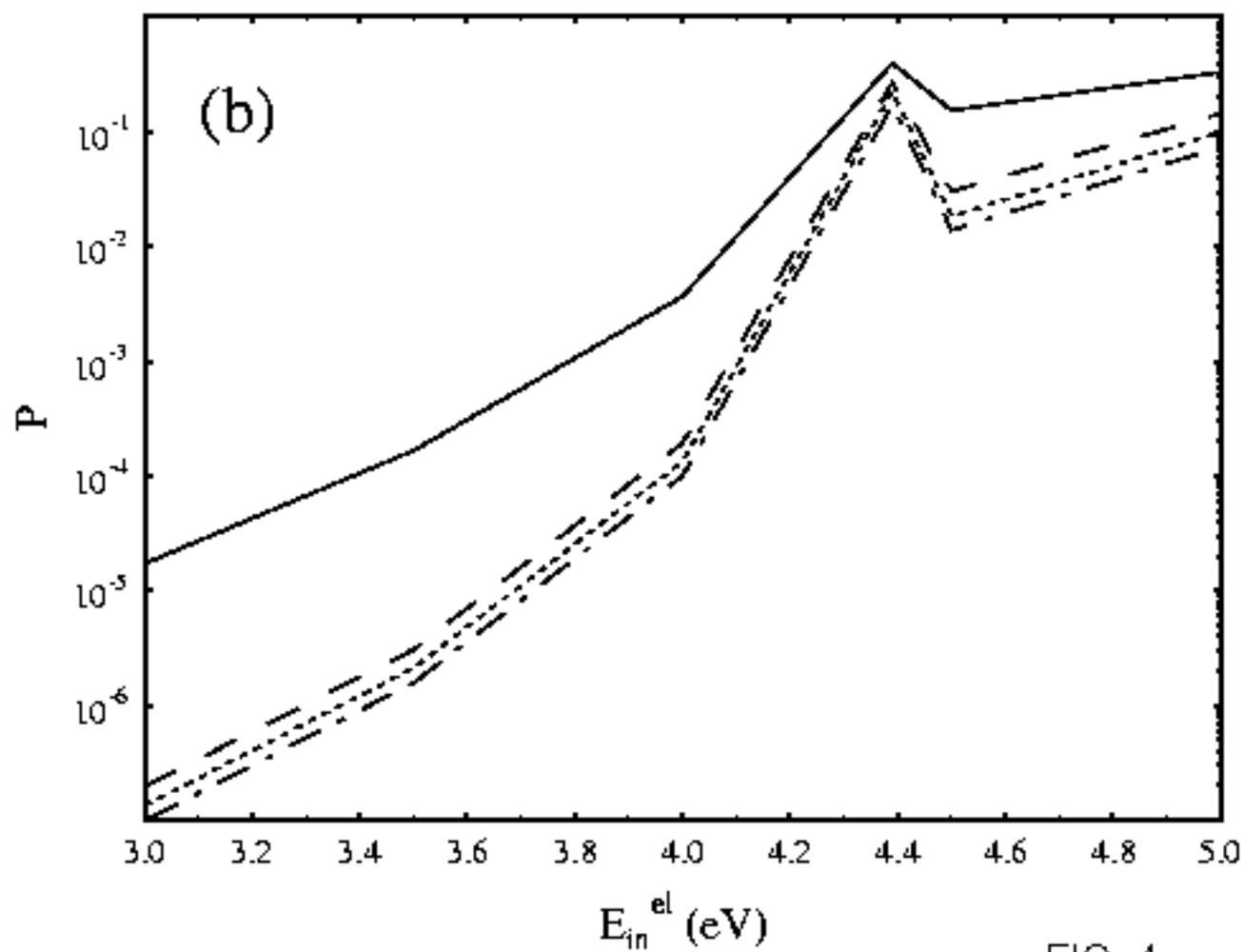

FIG. 4.

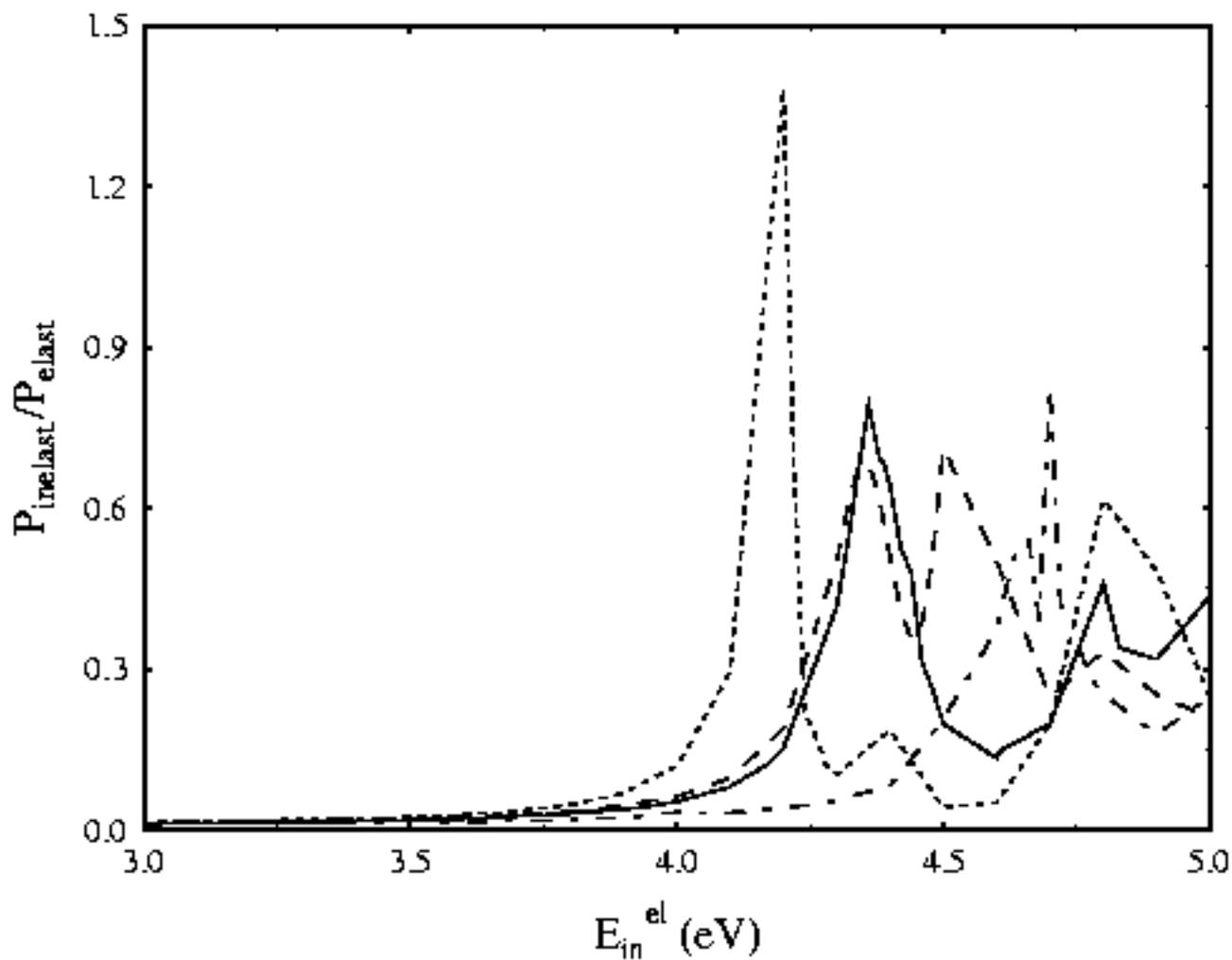

FIG. 5.

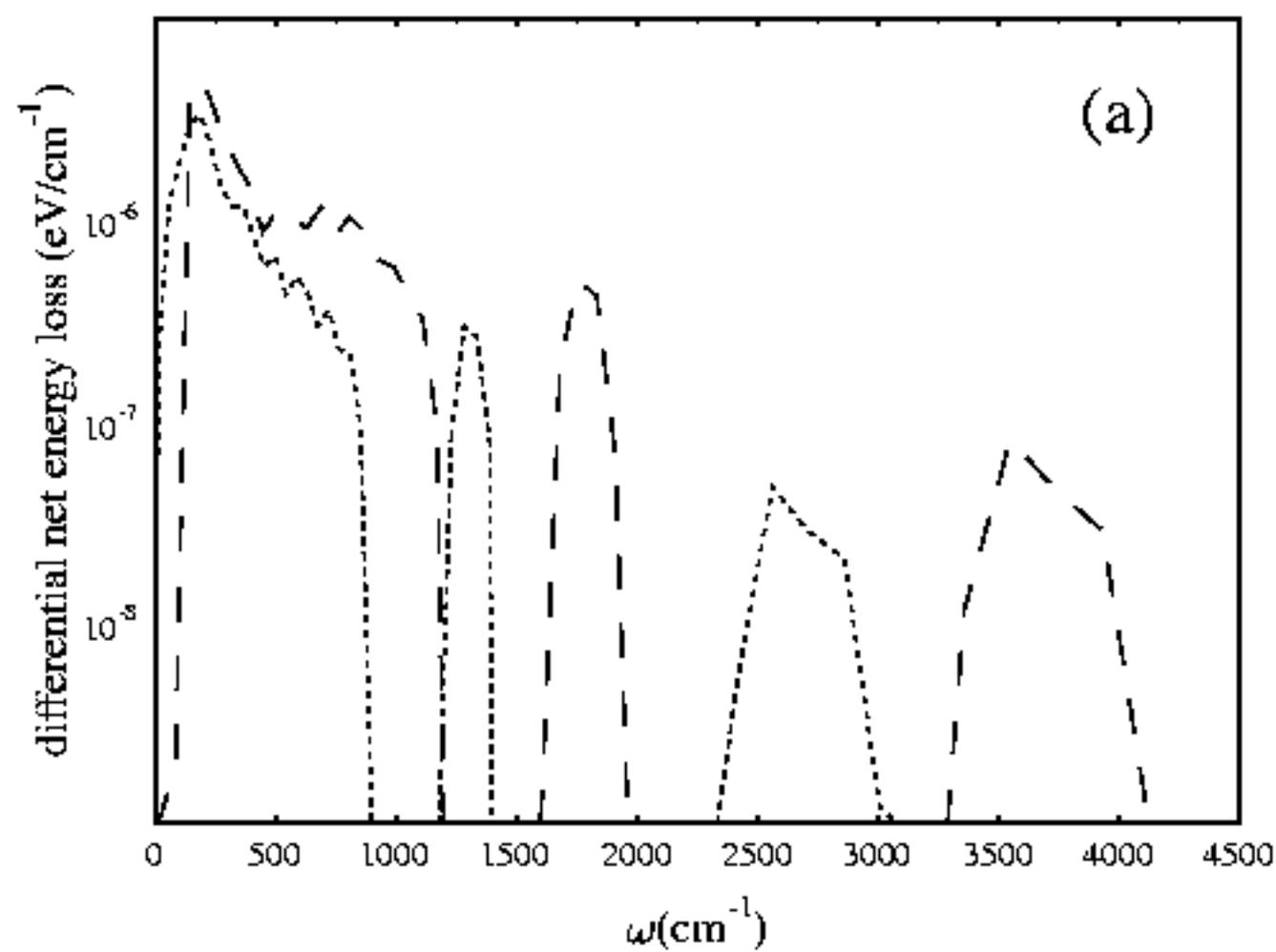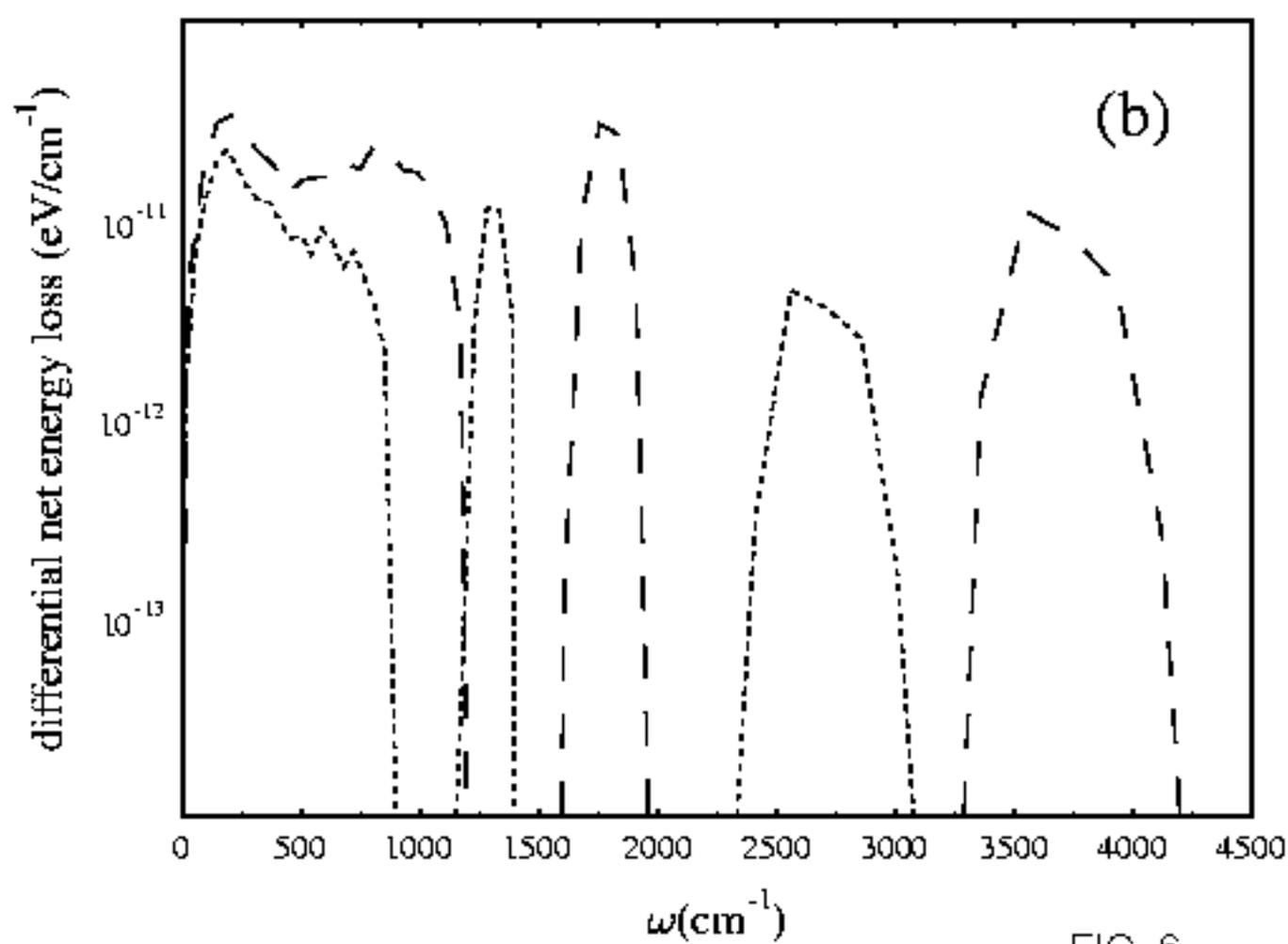

FIG. 6

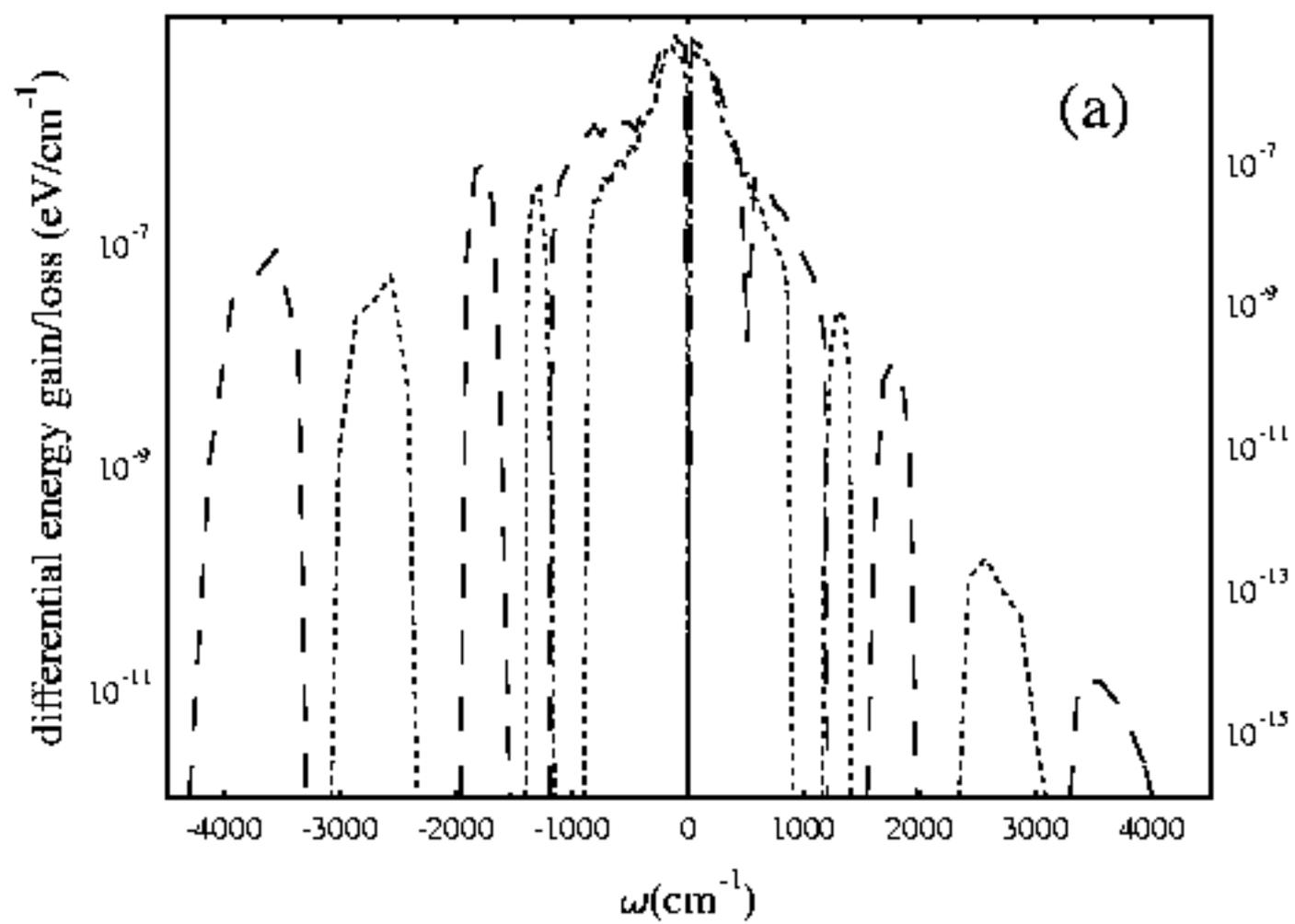

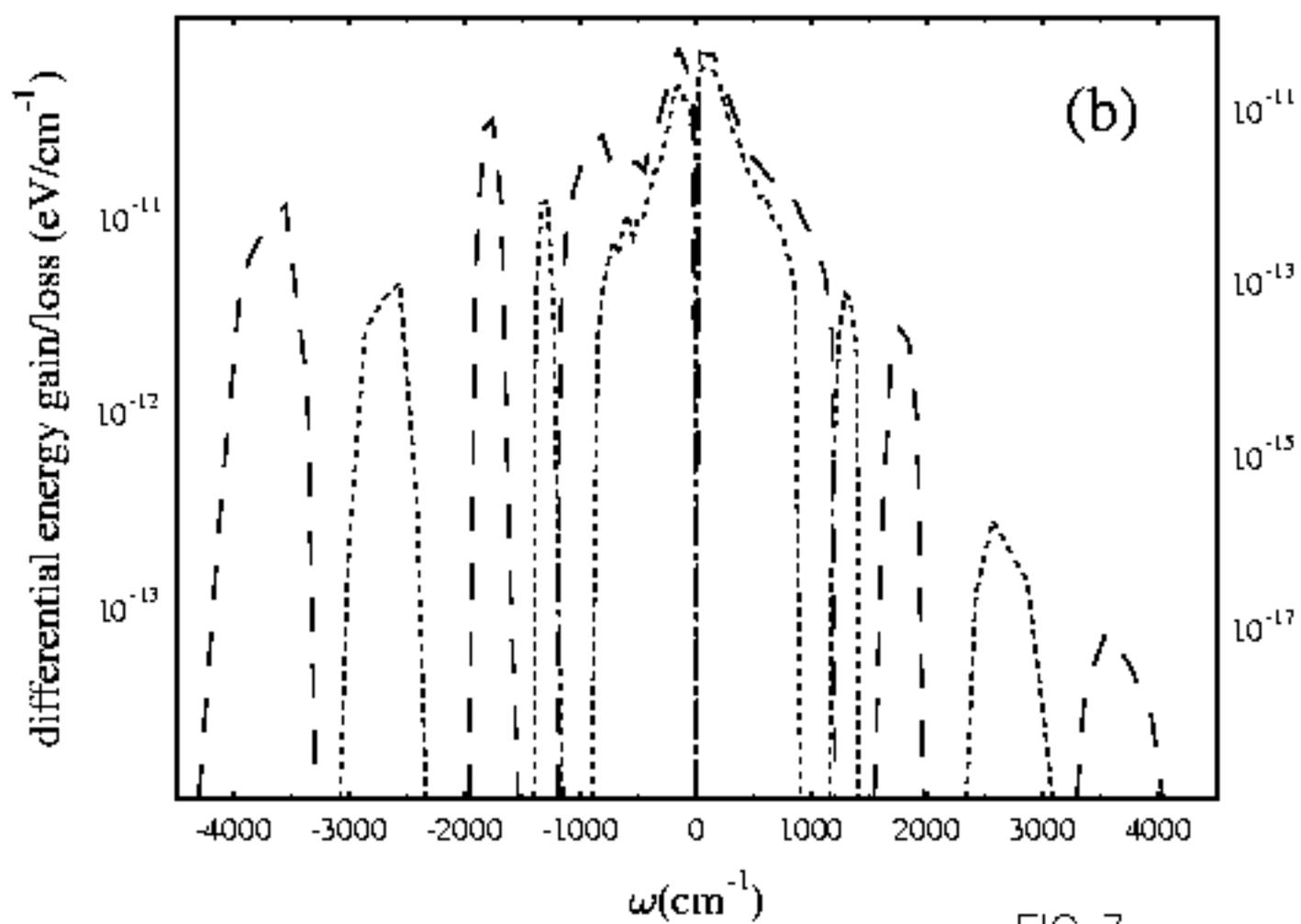

FIG. 7.

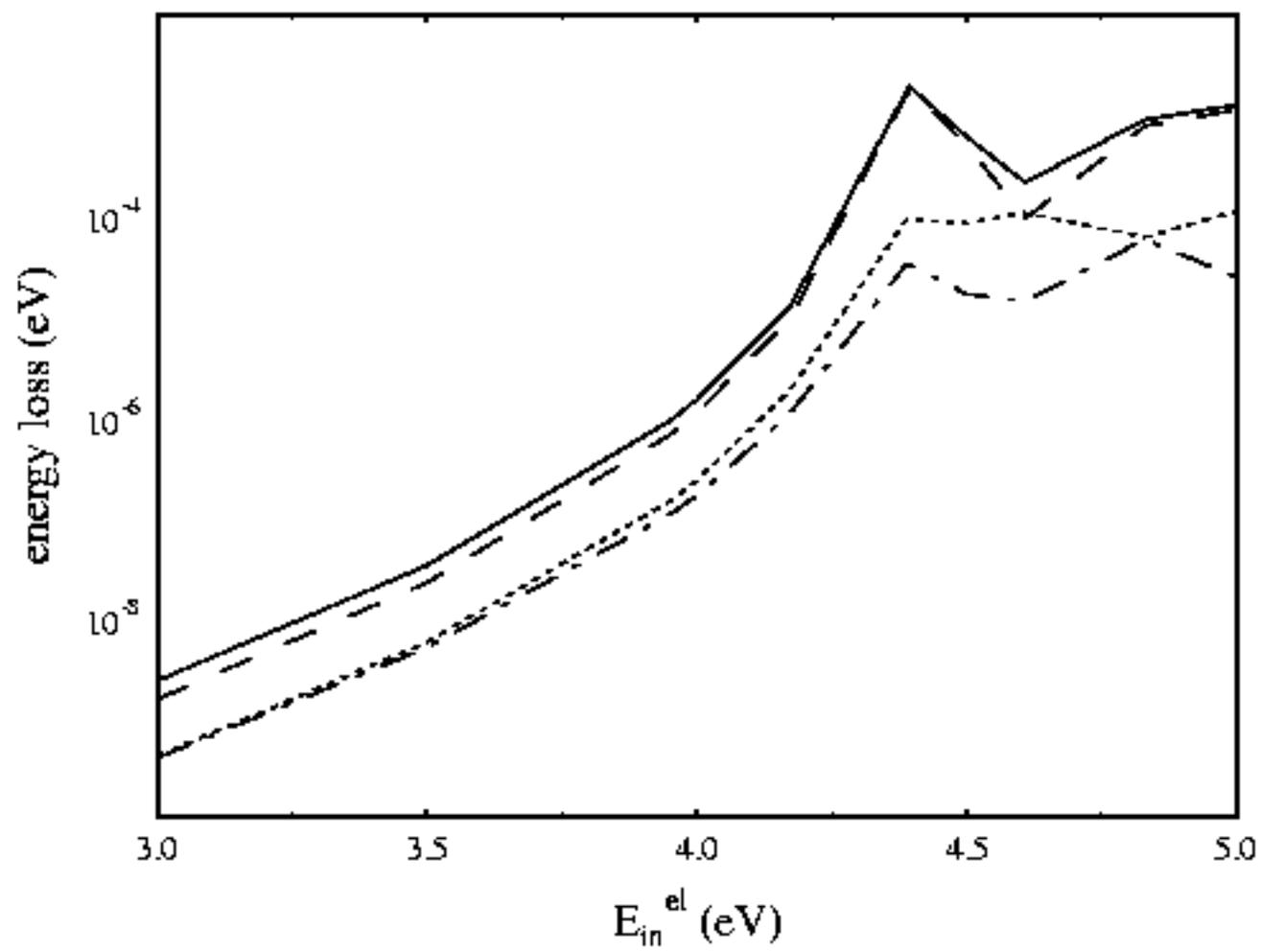

FIG. 8.